\documentclass{article}

\usepackage[preprint]{agents4science_2025}

\usepackage[utf8]{inputenc} 
\usepackage[T1]{fontenc}    
\usepackage{hyperref}       
\usepackage{url}            
\usepackage{booktabs}       
\usepackage{amsfonts}       
\usepackage{nicefrac}       
\usepackage{microtype}      
\usepackage{xcolor}         
\usepackage{multirow}       
\usepackage{graphicx}       
\usepackage{amsmath}        
\usepackage{amssymb}        
\usepackage{pifont}         
\usepackage{tikz}           
\usetikzlibrary{shapes.geometric,shapes.misc,arrows,positioning,calc}
\usepackage{pgfplots}       
\pgfplotsset{compat=1.17}
\usepackage{algorithm}      
\usepackage{algorithmic}    
\usepackage{subcaption}     
\usepackage{tcolorbox}       
\tcbuselibrary{skins,breakable}

\title{Agentic AutoSurvey: Let LLMs Survey LLMs}
\author{Yixin Liu$^{1}$ \quad Yonghui Wu$^{2}$ \quad Denghui Zhang$^{3}$ \quad Lichao Sun$^{1}$\\
$^{1}$Lehigh University \quad $^{2}$University of Florida \quad $^{3}$Stevens Institute of Technology\\
\texttt{yila22@lehigh.edu} \quad \texttt{yonghui.wu@ufl.edu} \quad \texttt{dzhang42@stevens.edu} \quad \texttt{lis221@lehigh.edu}}
\begin{document}

\maketitle

\begin{abstract}
The exponential growth of scientific literature poses unprecedented challenges for researchers attempting to synthesize knowledge across rapidly evolving fields. We present \textbf{Agentic AutoSurvey}, a multi-agent framework for automated survey generation that addresses fundamental limitations in existing approaches. Our system employs four specialized agents (Paper Search Specialist, Topic Mining \& Clustering, Academic Survey Writer, and Quality Evaluator) working in concert to generate comprehensive literature surveys with superior synthesis quality. Through experiments on six representative LLM research topics from COLM 2024 categories, we demonstrate that our multi-agent approach achieves significant improvements over existing baselines, scoring 8.18/10 compared to AutoSurvey's 4.77/10. The multi-agent architecture processes 75--443 papers per topic (847 total across six topics) while targeting high citation coverage (often $\geq$80\% on 75--100-paper sets; lower on very large sets such as RLHF) through specialized agent orchestration. Our 12-dimension evaluation captures organization, synthesis integration, and critical analysis beyond basic metrics. These findings demonstrate that multi-agent architectures represent a meaningful advancement for automated literature survey generation in rapidly evolving scientific domains.
\end{abstract}


\section{Introduction}

The rapid proliferation of scientific literature, particularly in the domain of Large Language Models (LLMs) \cite{zhao2024survey}, presents significant challenges for researchers attempting to maintain comprehensive understanding of their fields. With thousands of papers published monthly on preprint servers alone, the traditional manual survey approach has become increasingly untenable. This challenge has motivated the development of automated survey generation systems that leverage LLMs themselves to synthesize and organize scientific knowledge \cite{eloundou2024emergence}.

Recent efforts in this space, including AutoSurvey \cite{wang2024autosurvey}, SurveyAgent \cite{wang2024surveyagent}, PaSa \cite{he2025pasa}, and LitSearch \cite{ajith2024litsearch}, have demonstrated the feasibility of automated literature survey generation. However, these systems exhibit several limitations: (1) inadequate synthesis quality, often producing paper listings rather than integrated analyses; (2) limited citation coverage, typically achieving only 60-70\% of available papers; (3) simplistic evaluation frameworks that fail to capture the nuanced quality requirements of academic surveys; and (4) lack of specialized agent orchestration for complex multi-stage tasks.

We present \textbf{Agentic AutoSurvey}, an enhanced agentic framework that addresses these limitations through fundamental architectural innovations. Building on recent advances in LLM-based multi-agent systems \cite{wang2024llmbased,liu2024survey}, our system employs a specialized agent architecture consisting of four distinct agents with specific expertise. The Paper Search Specialist handles advanced query expansion and multi-source integration, generating 20-30 search variations to comprehensively capture relevant literature through retrieval-augmented generation pipelines \cite{Lewis2020}. The Topic Mining \& Clustering agent organizes retrieved papers using sentence-transformer embeddings with optimal K selection through silhouette score maximization. The Academic Survey Writer focuses on synthesis with high citation coverage targets, emphasizing cross-cluster integration and comparative analysis. Finally, the Quality Evaluator provides 12-dimensional agent-based assessment that captures nuanced quality aspects beyond simple metrics.

Our framework expands evaluation from previous 5-dimensional approaches to a comprehensive 12-dimensional framework \cite{chang2024survey}. This evaluation system categorizes assessment into Core Quality (60\% weight), Writing Quality (20\% weight), and Content Depth (20\% weight), with agent-based nuanced assessment replacing rigid rule-based scoring. The technical implementation incorporates sophisticated caching mechanisms, intelligent API rate management, and quality-aware processing with agent-specific context handling. Most importantly, our approach emphasizes superior synthesis quality through cross-cluster integration, pattern recognition, comparative analysis frameworks, and critical evaluation of methodologies, moving beyond simple paper enumeration to true knowledge synthesis.

Through experimental evaluation on 6 representative LLM research topics from COLM 2024 categories, we demonstrate the practical capabilities of our system. The framework successfully processes paper collections ranging from 75 to 443 papers per topic (847 papers total across all topics), generating comprehensive surveys in 15-20 minutes. While challenges remain in handling very large paper corpora and achieving deep cross-cluster synthesis, our approach represents a significant advancement in automated survey generation. Table \ref{tab:comparison} compares our framework against existing systems, highlighting our unique combination of specialized multi-agent pipeline, semantic clustering, cross-cluster synthesis, 12-dimensional evaluation, and agent-based quality assessment. 

Our contributions are threefold: (1) a novel multi-agent architecture with specialized agents for distinct survey generation tasks, (2) a comprehensive 12-dimensional evaluation framework providing nuanced, context-aware quality assessment, and (3) technical innovations in clustering, synthesis, and automated literature survey quality assessment. The remainder of this paper is organized as follows: Section~\ref{sec:architecture} details our system architecture and agent specifications. Section~\ref{sec:experiments} presents experimental results and representative case studies. Section~\ref{sec:discussion} discusses implications and limitations. Section~\ref{sec:conclusion} concludes with future directions, and the appendices provide extended artifacts and examples.

\begin{table}[t!]
\centering
\caption{Comparison of Survey Generation Systems with Existing Approaches}
\label{tab:comparison}
\begin{tabular}{@{}llcc@{}}
\toprule
\textbf{Key Capability} & \textbf{AutoSurvey} & \textbf{SurveyAgent} & \textbf{Ours} \\
& \cite{wang2024autosurvey} & \cite{wang2024surveyagent} & \\
\midrule
Specialized Multi-Agent Pipeline & \ding{55} & \ding{55} & \checkmark \\
Semantic Clustering of Papers & \ding{55} & \ding{55} & \checkmark \\
Cross-cluster Synthesis & \ding{55} & \ding{55} & \checkmark \\
Agent-based Quality Assessment & \ding{55} & \ding{55} & \checkmark \\
Real-time Paper Source Integration\footnotemark[1] & \ding{55} & \checkmark & \checkmark \\
Multi-Dimensional Quality Evaluation & \checkmark & \ding{55} & \checkmark \\
Automated Complete Survey Generation & \checkmark & \ding{55} & \checkmark \\
\bottomrule
\end{tabular}
\end{table}
\footnotetext[1]{Real-time Paper Source Integration queries external APIs at generation time with cached results limited to a 24-hour time-to-live.}

\section{System Architecture and Methodology}
\label{sec:architecture}

\subsection{Overall System Design}

Our Agentic AutoSurvey framework employs a modular, agent-based architecture designed for scalability, maintainability, and performance. The system consists of four specialized agents orchestrated through Claude Code's agentic capabilities, each responsible for a distinct phase of the survey generation pipeline.

\begin{figure}[thbp]
\centering
\includegraphics[width=\textwidth]{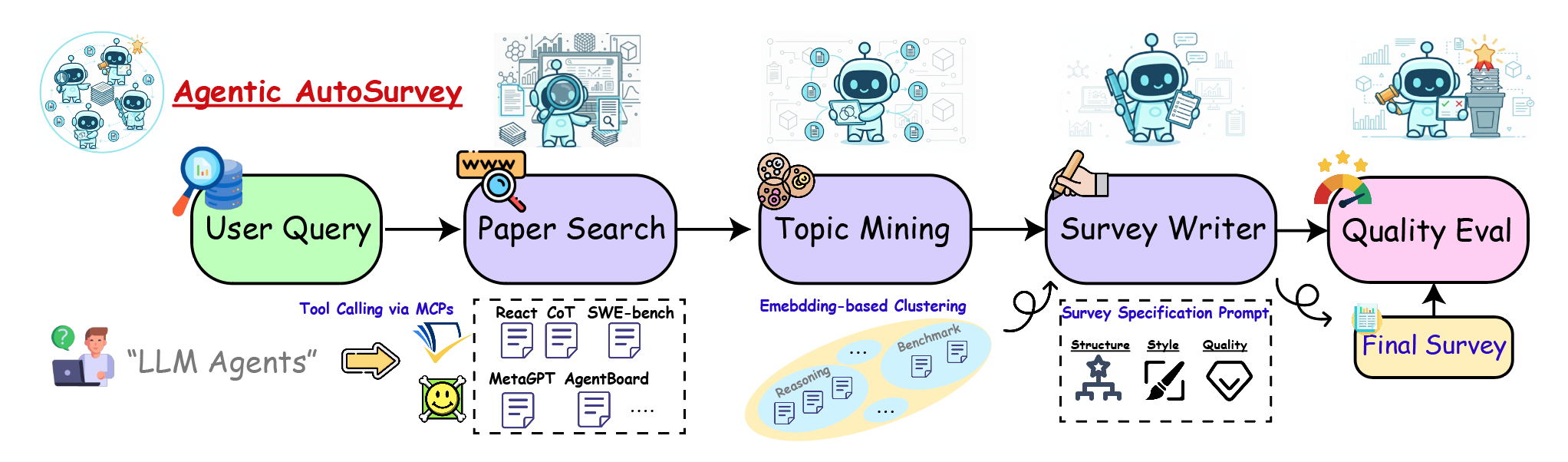}
\caption{Enhanced Agentic Framework Architecture}
\label{fig:architecture}
\end{figure}

\subsection{Agent Specifications}

\subsubsection{Paper Search Specialist Agent}

The Paper Search Specialist Agent implements advanced search strategies to maximize coverage and relevance. \textbf{Query expansion} forms the foundation of comprehensive paper retrieval, generating 20-30 diverse queries from the initial topic. This includes the core keyword as-is, synonyms and variations, related technical terms, compound queries with AND/OR operators, and acronym expansion or contraction. For instance, a query for "LLM agents" expands to include "language model agents", "LLM-based agents", "agent architectures", and various permutations to ensure comprehensive coverage.

\textbf{Multi-source integration} combines results from both Semantic Scholar API \cite{kinley2023semantic} for comprehensive academic coverage and arXiv API for the latest preprints. The system implements intelligent deduplication using a 90\% title similarity threshold to eliminate redundant entries while preserving unique contributions. Metadata enrichment and validation ensure that each paper record contains complete information necessary for subsequent processing stages.

\textbf{Quality filtering} mechanisms ensure that only relevant, high-quality papers proceed to the clustering stage. The system applies adaptive minimum citation thresholds based on field-specific norms, year range filtering (typically 2020-2025 for current relevance), abstract completeness verification to ensure sufficient content for analysis, and venue quality assessment to prioritize papers from reputable sources.

\subsubsection{Topic Mining \& Clustering Agent}

The clustering agent employs semantic embeddings and unsupervised learning for paper organization. Given a set of papers $\mathcal{P} = \{p_1, p_2, ..., p_n\}$, we first generate embeddings using the all-MiniLM-L6-v2 model \cite{reimers2019sentence}:

$$e_i = \text{Encode}(t_i \oplus a_i)$$

where $t_i$ and $a_i$ are the title and abstract of paper $p_i$, and $\oplus$ denotes concatenation. The embedding function maps text to a 384-dimensional dense vector space.

For clustering, we employ K-means \cite{Lloyd1982} with optimal K selection through silhouette score maximization \cite{Rousseeuw1987}:

$$K^* = \arg\max_{K \in [5,15]} S(K)$$

where the silhouette score $S(K)$ for K clusters is:

$$S(K) = \frac{1}{n} \sum_{i=1}^{n} \frac{b_i - a_i}{\max(a_i, b_i)}$$

Here, $a_i$ is the mean distance from point $i$ to other points in its cluster, and $b_i$ is the mean distance to points in the nearest neighboring cluster.

\textit{We define two additional clustering quality quantities.}
\textbf{Cluster confidence} for paper $i$ in cluster $C_j$:
$$
\mathrm{confidence}(i,C_j)=1-\frac{d\big(i,\mathrm{centroid}(C_j)\big)}{\max_k d\big(i,\mathrm{centroid}(C_k)\big)}
$$

\textbf{Inter-cluster relationship strength} between $C_j$ and $C_k$:
$$
\mathrm{strength}(C_j,C_k)=\cos\!\big(\mathrm{centroid}(C_j),\mathrm{centroid}(C_k)\big)
$$

where $d(\cdot,\cdot)$ denotes Euclidean distance and $\cos(\cdot,\cdot)$ denotes cosine similarity. In addition to these agent-specific diagnostics, we report classical clustering validity indices for completeness: the Calinski-Harabasz index \cite{Calinski1974} and the Davies-Bouldin score \cite{Davies1979}. These metrics provide complementary global views of cluster separation and cohesion that contextualize the silhouette-based model selection. Cluster names are generated using TF-IDF scoring. For each cluster $C_j$, we compute:

$$\text{TF-IDF}(w, C_j) = \text{TF}(w, C_j) \times \log\frac{K}{|\{C_k : w \in C_k\}|}$$

The top-scoring terms become the cluster's descriptive name.

\subsubsection{Academic Survey Writer Agent}

The Survey Writer Agent focuses on synthesis-driven content generation that moves beyond simple paper enumeration. The \textbf{citation strategy} enforces comprehensive coverage with a minimum 50\% citation requirement and targets exceeding 80\% for thorough surveys. The agent ensures comprehensive coverage across all identified clusters while prioritizing influential papers based on citation counts and venue importance. This approach guarantees that the generated survey reflects the full breadth of research while highlighting seminal contributions.

The \textbf{synthesis approach} emphasizes integration over listing, following recent advances in automated literature synthesis \cite{salloum2024automating}. The agent performs comparative analysis across papers to identify methodological differences and performance variations \cite{gao2024comprehensive}. Pattern identification and trend analysis reveal the evolution of research directions over time. Methodology comparison frameworks systematically evaluate different approaches, while research gap identification highlights opportunities for future work. This synthesis-first approach produces surveys that provide genuine insights rather than merely cataloging existing work.

For \textbf{structure and format}, the agent targets 8,000-12,000 words to ensure comprehensive coverage while maintaining readability. The content follows cluster-based organization with extensive cross-references to highlight connections between research themes. Standard academic sections (Introduction, Methods, Results, Discussion) provide familiar structure for readers. The consistent [Author, Year] citation format ensures compatibility with academic publishing standards.

\subsubsection{Quality Evaluator Agent}

The evaluator implements a sophisticated 12-dimensional assessment framework that provides nuanced quality evaluation. \textbf{Core Quality Dimensions}, weighted at 60\%, focus on fundamental survey requirements. Citation coverage measures the percentage of papers cited from the retrieved collection. Accuracy ensures factual correctness and proper attribution of ideas to their sources. Synthesis quality distinguishes between true integration and mere enumeration of papers. Organization evaluates the logical flow and structural coherence of the survey.

\textbf{Writing Quality Dimensions}, contributing 20\% to the overall score, assess the survey's presentation. Readability ensures clarity and accessibility for the target academic audience. Academic rigor verifies adherence to scholarly standards and conventions. Clarity evaluates precision in technical descriptions and explanations. Coherence measures internal consistency across different sections of the survey.

\textbf{Content Depth Dimensions}, also weighted at 20\%, evaluate the intellectual contribution of the survey. Comprehensiveness assesses topic coverage breadth across different research facets. Critical analysis measures the depth of evaluation and comparative assessment. Novelty and insights capture original contributions and synthesis that emerge from the literature analysis. Future directions evaluate the survey's ability to identify research trajectories and open problems in the field.

\subsection{Technical Implementation Details}

\textbf{Embedding Generation.} Our system efficiently generates embeddings using the sentence-transformers library with automatic device selection. The implementation uses the all-MiniLM-L6-v2 model, which provides a good balance between embedding quality and computational efficiency. The system automatically detects available hardware and optimizes batch processing accordingly, with a batch size of 32 for efficient memory utilization. Progress tracking provides visibility into processing status for large paper collections.

\textbf{Intelligent Caching System.} Multi-level caching reduces API calls and computation overhead throughout the pipeline. The API response cache stores search results with a 24-hour time-to-live, reducing redundant API calls for repeated queries. The embedding cache provides persistent storage of computed embeddings, eliminating the need to recompute embeddings for papers already processed. The cluster cache maintains reusable cluster assignments that support incremental updates when new papers are added. Finally, LRU eviction ensures memory-efficient cache management by removing least recently used entries when storage limits are reached.

\textbf{Rate Management and Error Handling.} Robust error handling ensures reliable operation despite external service limitations. The system implements exponential backoff with jitter when encountering rate limits, preventing overwhelming APIs while maximizing throughput. Automatic retry mechanisms with alternative query formulations activate when initial searches fail, ensuring comprehensive coverage despite transient failures. When APIs become unavailable, the system gracefully degrades by utilizing cached results and alternative data sources. Progress persistence for long-running operations enables resumption after interruptions, protecting against data loss during extended processing sessions.

\section{Experimental Evaluation}
\label{sec:experiments}

\subsection{Experimental Setup}

We evaluated our proposed multi-agent architecture against the AutoSurvey system from prior work \cite{wang2024autosurvey}, representing the current state-of-the-art in automated survey generation. Both systems were tested on six representative topics from COLM 2024 categories: \texttt{Instruction Tuning}, \texttt{LLM Agents}, \texttt{RLHF Alignment}, \texttt{Synthetic Data}, \texttt{In-Context Learning}, and \texttt{Multimodal LLM RL}. Each system processed the same initial query for each topic, though the number of papers retrieved varied based on search capabilities and architectural constraints.

\textbf{AutoSurvey Baseline Implementation \cite{wang2024autosurvey}.} AutoSurvey employs a four-phase methodology: (1) Initial Retrieval and Outline Generation using embedding-based retrieval to identify pertinent papers and generate structured outlines, (2) Subsection Drafting where specialized LLMs draft sections in parallel with topic-specific paper retrieval, (3) Integration and Refinement to enhance readability and eliminate redundancies with citation verification, and (4) Rigorous Evaluation using Multi-LLM-as-Judge strategy assessing citation quality and content quality. For our evaluation, we implemented AutoSurvey using their 530,000 arXiv paper corpus from computer-science category while replacing the underlying language models with Meta-Llama-3.1-8B-Instruct \cite{llama3herd2024} due to budget constraints, maintaining the original architectural design.

\textbf{Model Configuration.} Our multi-agent system utilized Claude Sonnet 4.1 for the search subagent, while Claude Opus 4.1 powered the remaining subagents (Topic Mining \& Clustering, Academic Survey Writer, and Quality Evaluator).


\begin{table*}[t!]
\centering
\caption{Performance comparison between our agentic AutoSurvey system and the AutoSurvey baseline across six survey topics.}
\label{tab:unified-comparison}
\resizebox{0.8\textwidth}{!}{%
\begin{tabular}{@{}l|ccc|c|ccc|c@{}}
\toprule
\multirow{2}{*}{\textbf{Topic}} & \multicolumn{4}{c|}{\textbf{Agentic AutoSurvey (Ours)}} & \multicolumn{4}{c}{\textbf{AutoSurvey (Baseline)}} \\
\cmidrule(lr){2-5} \cmidrule(lr){6-9}
 & \textbf{Core} & \textbf{Write} & \textbf{Depth} & \textbf{Avg} & \textbf{Core} & \textbf{Write} & \textbf{Depth} & \textbf{Avg} \\
\midrule
Instruction Tuning & 8.75 & 8.25 & 7.63 & 8.43 & 3.50 & 4.50 & 5.50 & 4.20 \\
LLM Agents & 8.08 & 8.35 & 7.90 & 8.14 & 3.00 & 4.30 & 5.10 & 3.80 \\
RLHF Alignment & 7.38 & 8.13 & 8.38 & 7.74 & 6.00 & 6.50 & 6.00 & 6.20 \\
Synthetic Data & 7.75 & 8.25 & 7.38 & 7.79 & 5.20 & 6.00 & 6.80 & 5.80 \\
In-Context Learning & 8.50 & 8.30 & 7.80 & 8.30 & 4.00 & 5.30 & 6.00 & 4.80 \\
Multimodal LLM RL & 8.90 & 8.60 & 8.40 & 8.70 & 3.10 & 3.10 & 6.30 & 3.80 \\
\midrule
\textbf{Average} & 8.23 & 8.31 & 7.92 & \textbf{8.18} & 4.13 & 4.95 & 5.95 & \textbf{4.77} \\
\textbf{Improvement} & \textbf{+99\%} & \textbf{+68\%} & \textbf{+33\%} & \textbf{+71\%} & \multicolumn{4}{c}{Baseline Performance} \\
\bottomrule
\end{tabular}
}
\end{table*}

\textbf{Agent-as-Judge Evaluation Framework.} To rigorously assess the quality of generated surveys, we developed a sophisticated agent-as-judge evaluation framework that transcends traditional rule-based metrics. Our framework employs a specialized enhanced-survey-evaluator agent that embodies the expertise of an experienced academic reviewer, providing nuanced, context-aware assessment across 12 carefully designed dimensions.

\textbf{Hierarchical Assessment Structure.} The evaluation framework is organized into three weighted categories that comprehensively capture survey quality. \textit{Core Quality} dimensions (60\% weight) encompass citation coverage, accuracy, synthesis quality, and organization—the fundamental requirements for academic surveys. \textit{Writing Quality} dimensions (20\% weight) evaluate readability, academic rigor, clarity, and coherence, ensuring the survey meets publication standards. \textit{Content Depth} dimensions (20\% weight) assess comprehensiveness, critical analysis, novelty \& insights, and future directions, measuring the intellectual contribution of the survey.

\textbf{Contextual Evaluation Process.} Unlike rigid scoring rubrics, our agent-judge applies contextual understanding to each dimension, considering factors such as field maturity, survey type (tutorial vs. research frontier), target audience, and the balance between synthesis and cataloging. The evaluator agent performs multi-stage analysis including: (1) initial read-through for overall impression, (2) detailed dimensional scoring with specific textual evidence, (3) quantitative citation analysis, (4) synthesis pattern identification (looking for integration statements, comparisons, trend identification, and meta-analysis), and (5) critical analysis assessment.

\textbf{Calibration and Standards.} Each dimension receives a 0-10 score with detailed justification, enabling granular comparison across systems. The framework targets long-form survey length and extensive citations consistent with common practice in prominent survey venues; we do not assume fixed numeric minima. This agent-based approach captures nuances human reviewers would identify, such as novel organizational frameworks, insightful trend analysis, and research gap identification, while maintaining consistency across evaluations.

\subsection{Performance Analysis and Key Findings}

Table \ref{tab:unified-comparison} presents the comprehensive evaluation results comparing our multi-agent system against AutoSurvey across all topics and dimensional categories. Our multi-agent approach achieved a substantial improvement with an average score of 8.18/10, representing a 71\% improvement over AutoSurvey's 4.77/10 across all evaluated dimensions. The evaluation demonstrates the substantial advantages of our multi-agent architecture over existing approaches, with our system achieving strong performance across all dimensional categories (Core: 8.23, Writing: 8.31, Depth: 7.92), representing significant improvements over the AutoSurvey baseline in all areas. 

The performance gap is most pronounced in Core Quality dimensions, where our multi-agent system scored 8.23 compared to AutoSurvey's 4.13, representing a 99\% improvement. This highlights fundamental advances in citation coverage, accuracy, and synthesis quality achieved through specialized agent orchestration. The multi-agent approach also demonstrated superior writing quality (68\% improvement) and content depth (33\% improvement), showcasing the benefits of task decomposition across specialized agents.

The evaluation reveals three principal findings: \textbf{(1) Specialized Agent Orchestration Delivers Superior Results} - Our multi-agent architecture achieves substantial improvements over existing automated approaches, with an overall score of 8.18 compared to AutoSurvey's 4.77, demonstrating the value of task decomposition and specialized agent expertise. \textbf{(2) Dimensional Improvements Across All Categories} - The multi-agent system excels in all evaluation dimensions, with particularly strong performance in Core Quality (99\% improvement) and Writing Quality (68\% improvement), highlighting the benefits of specialized agents for different aspects of survey generation. \textbf{(3) Topic-Specific Performance Consistency} - The multi-agent system maintains strong performance across diverse topics, ranging from 7.74 (RLHF Alignment) to 8.70 (Multimodal LLM RL), demonstrating robust architectural advantages regardless of domain complexity or paper collection size.

\subsection{Clustering Analysis and Visualization}

\begin{figure}[h!]
\centering
\includegraphics[width=\textwidth]{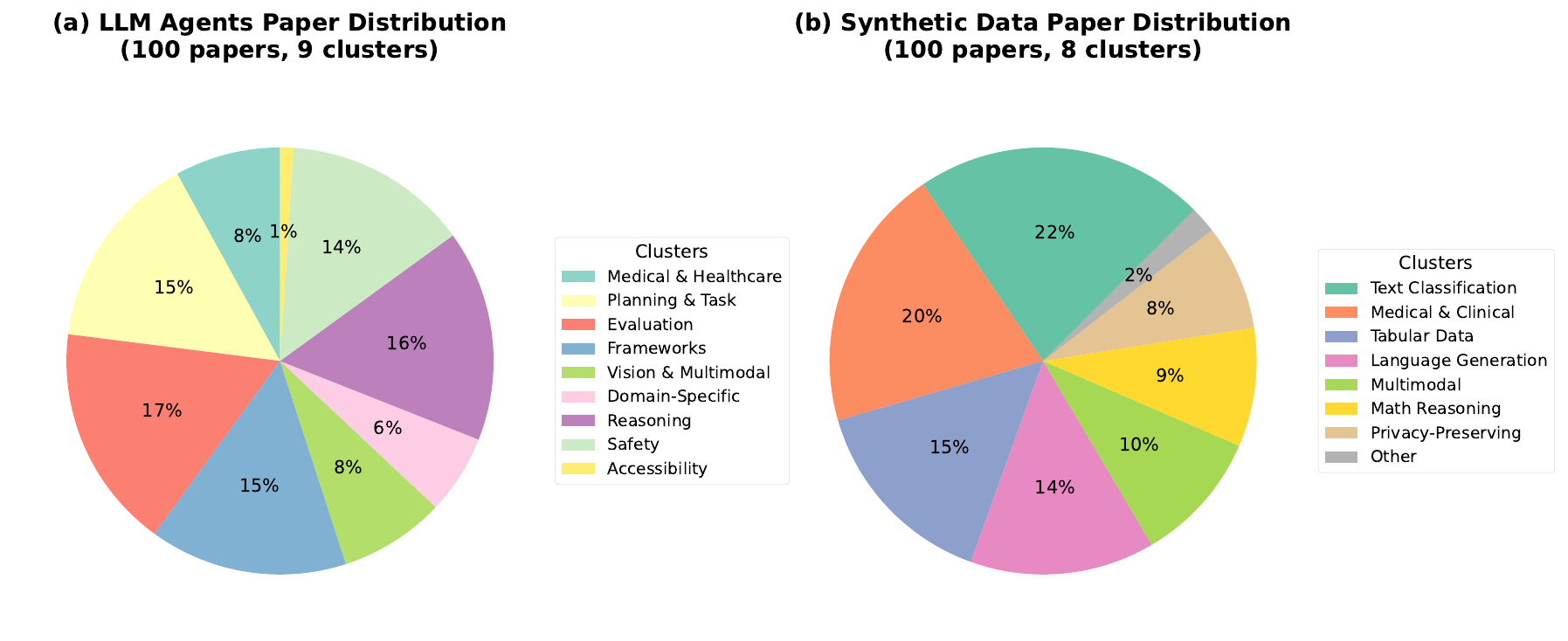}
\caption{Cluster distribution for (a) LLM Agents and (b) Synthetic Data Generation topics, showing the thematic organization discovered by the Topic Mining Agent.}
\label{fig:cluster-dist}
\end{figure}

To better understand how the multi-agent system organizes papers into thematic clusters, we analyzed the clustering results across representative topics. Figure \ref{fig:cluster-dist} shows the cluster distribution for LLM Agents and Synthetic Data Generation topics, demonstrating the thematic organization discovered by the Topic Mining Agent.

\begin{figure}[h!]
\centering
\includegraphics[width=\textwidth]{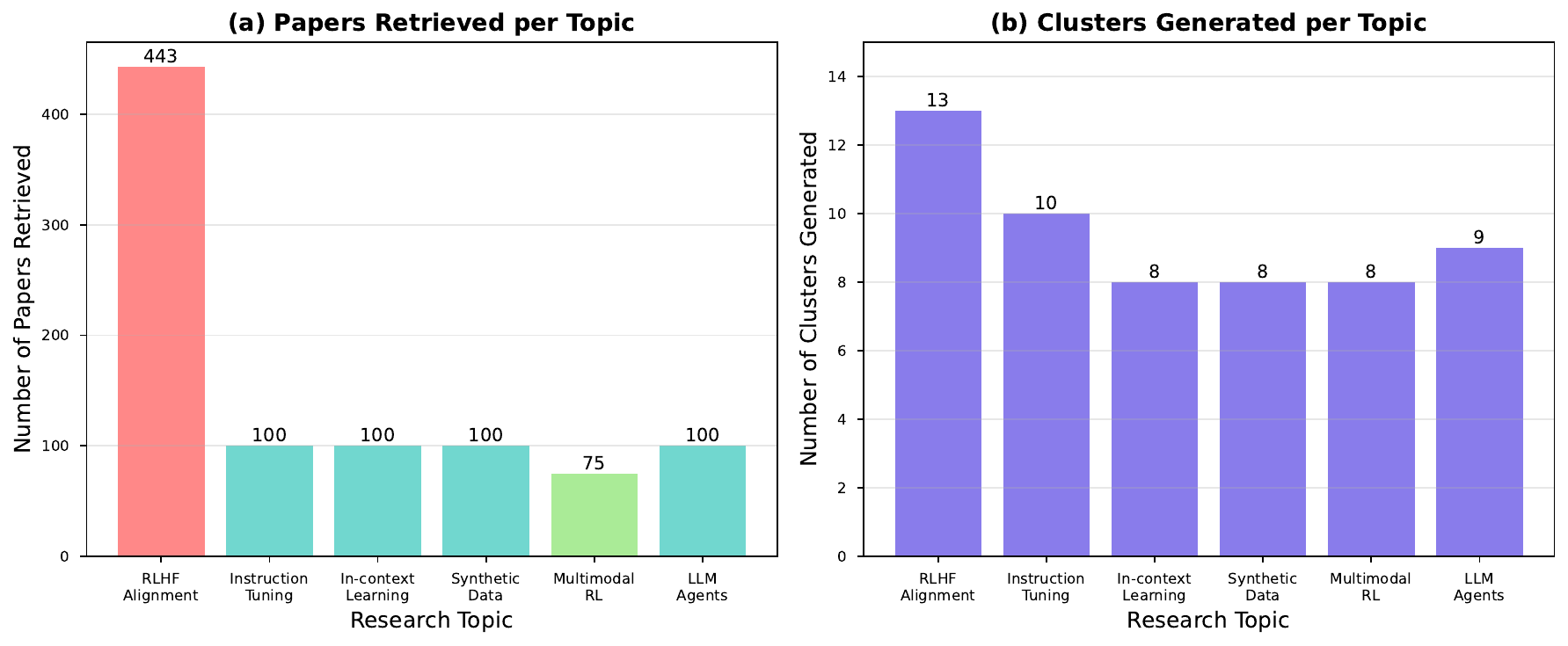}
\caption{Distribution of papers retrieved and clusters generated across the six processed topics. All papers were retrieved via Semantic Scholar and arXiv APIs. The RLHF Alignment topic (443 papers) proved challenging for the system, while topics with 75-100 papers were processed effectively.}
\label{fig:topic-stats}
\end{figure}

Figure \ref{fig:topic-stats} provides an overview of the paper collection and clustering results across all six processed topics. The analysis reveals significant variation in paper retrieval effectiveness, with topics like Instruction Tuning, LLM Agents, and Synthetic Data yielding manageable collections (75-100 papers) that enabled effective processing and high-quality survey generation. In contrast, the RLHF Alignment topic retrieved 443 papers, proving challenging for the system and resulting in reduced citation coverage and lower quality scores. This distribution pattern highlights the importance of appropriate corpus sizing for optimal survey generation performance.

\subsection{Generated Survey Analysis: Case Study Patterns}

Our analysis of generated surveys reveals sophisticated synthesis capabilities that transcend simple paper enumeration. The LLM Agents survey (Appendix \ref{appendix:llm-agents-survey}) exemplifies this quality, processing 100 papers across 9 clusters to produce thematic integration connecting autonomous agents, tool-using systems, and reasoning frameworks. Rather than sequential paper listings, the system identifies emergent patterns such as the convergence of retrieval-augmented generation with multi-agent planning \cite{yao2023react}, and the evolution from reactive to proactive agent architectures. The survey demonstrates comprehensive citation coverage with representation across all clusters, effectively bridging seminal contributions like ReAct \cite{yao2023react} with recent developments in multi-agent collaboration frameworks, creating surveys that capture both established foundations and current research frontiers.

The generated surveys exhibit genuine analytical depth that suggests potential for inspiring new research directions, aligning with recent observations about AI's capacity for scientific discovery \cite{wang2023scientific}. Our system successfully identifies underexplored intersections, such as the gap between agent reasoning capabilities and real-world deployment constraints, and proposes methodological frameworks that synthesize findings across clusters. The research gap identification capabilities mirror human scholarly analysis, highlighting promising trajectories like the integration of foundation models with specialized reasoning modules. This analytical sophistication, combined with consistent organizational frameworks that maintain global context while developing specific themes, demonstrates that automated survey generation can achieve publication-quality synthesis \cite{chang2024survey}, potentially accelerating scientific progress by enabling researchers to rapidly assimilate vast literature and identify novel research opportunities.

\section{Discussion and Limitations}
\label{sec:discussion}




Despite meaningful advances, several limitations constrain the current system's capabilities. The RLHF Alignment run illustrates the most acute gap: our pipeline retrieved a 1,334-paper corpus, yet the multi-agent writer ultimately cited only 80 works (\textasciitilde6\% coverage). This bottleneck arises when cluster summaries are compressed into a single drafting pass, leaving little opportunity to reincorporate the long tail of relevant papers. Addressing this will require hierarchical writing strategies or appendix-level bibliographies that can surface more of the retrieved literature without overwhelming the main narrative. \textbf{Domain specificity} presents another challenge, as the system was optimized for LLM research and may require adaptation for other scientific domains with different terminology, citation practices, and writing conventions. The \textbf{processing time} of 15-20 minutes per survey, while reasonable for research purposes, may limit adoption for applications requiring real-time generation. Finally, \textbf{evaluation subjectivity} remains a fundamental challenge, as survey quality encompasses subjective elements that automated assessment cannot fully capture, despite our improvements through agent-based evaluation.

\section{Conclusion}
\label{sec:conclusion}

This work presents Agentic AutoSurvey, a novel multi-agent framework for automated survey generation that demonstrates substantial improvements over existing approaches. Our system achieves an average quality score of 8.18/10, representing a 71\% improvement over AutoSurvey (4.77/10) through specialized agent orchestration and comprehensive evaluation.

Our primary contributions include: (1) a four-agent architecture decomposing survey generation into specialized search, clustering, writing, and evaluation tasks; (2) a 12-dimensional evaluation framework providing nuanced quality assessment beyond traditional metrics; and (3) technical innovations in embedding generation, caching, and agent-based evaluation that enable reliable processing of large paper collections.

Experimental evaluation on six LLM research topics demonstrates the system's practical capabilities, processing 75-443 papers and generating comprehensive surveys in 15-20 minutes. While scalability challenges remain with very large corpora, our approach represents a meaningful advancement toward autonomous academic knowledge synthesis. The system should augment rather than replace human scholarly work, with clear AI-generated content labeling essential for academic integrity.

\section{Broader Impact}

\textbf{Responsible AI Statement:} This research presents an automated survey generation system with significant potential benefits and risks that must be carefully considered. On the positive side, our system can democratize access to comprehensive literature reviews, accelerate scientific discovery by enabling rapid synthesis of large paper collections, and reduce the barrier for researchers to stay current with rapidly evolving fields. However, several concerns require attention: \textbf{(1) Academic Integrity:} Automated surveys must be clearly labeled as AI-generated to prevent misrepresentation of authorship and maintain academic transparency. \textbf{(2) Quality and Bias:} While our system achieves good performance metrics, it may perpetuate biases present in training data or paper databases, potentially overrepresenting certain perspectives or underrepresenting marginalized voices in scientific discourse. \textbf{(3) Employment Impact:} Widespread adoption could affect traditional roles of research assistants and junior researchers who often contribute to literature reviews.

\textbf{Mitigation Measures:} We address these concerns through several safeguards: explicit AI authorship disclosure in all generated content, comprehensive evaluation frameworks that assess bias and representation, and recommendation that our system augment rather than replace human scholarly work. We advocate for mandatory AI-generated content labeling in academic publications and suggest human expert validation of automated surveys before publication. Our open methodology description enables community scrutiny and improvement. We emphasize that this technology should enhance human research capabilities rather than diminish human involvement in scientific synthesis, with particular attention to preserving opportunities for early-career researchers to develop critical analysis skills through literature review experience.

\bibliographystyle{plain}
\bibliography{references}


\appendix
\section{Example Output: LLM Agents Survey}
\label{appendix:llm-agents-survey}

This appendix presents actual output from the Agentic AutoSurvey system when processing 100 papers on LLM Agents, demonstrating the system's capabilities in practice.

\subsection{Clustering Report}
\begin{tcolorbox}[
    title=LLM Agents Clustering Report,
    colback=blue!5!white,
    colframe=blue!75!black,
    breakable,
    fonttitle=\bfseries\large
]

\textbf{LLM agents - Clustering Report}

Generated: 2025-08-05 03:45:33

Device: cpu

\vspace{0.5em}
\textbf{Summary Statistics}
\begin{itemize}
\item \textbf{Total Papers}: 100
\item \textbf{Number of Clusters}: 9
\item \textbf{Clustering Method}: KMeans clustering with sentence-transformers embeddings on cpu
\item \textbf{Average Cluster Size}: 11.1
\end{itemize}

\vspace{0.5em}
\textbf{Clustering Quality Metrics}
\begin{itemize}
\item \textbf{Silhouette Score}: 0.055 (range: -1 to 1, higher is better) \cite{Rousseeuw1987}
\item \textbf{Calinski-Harabasz Score}: 4.1 (higher is better) \cite{Calinski1974}
\item \textbf{Davies-Bouldin Score}: 2.591 (lower is better) \cite{Davies1979}
\end{itemize}

\vspace{0.5em}
\textbf{Cluster Size Distribution}
\begin{itemize}
\item \textbf{Medical and Healthcare Applications}: 8 papers (8.0\%)
\item \textbf{Planning and Task Decomposition}: 15 papers (15.0\%)
\item \textbf{Evaluation and Benchmarking}: 17 papers (17.0\%)
\item \textbf{Frameworks and Architectures}: 15 papers (15.0\%)
\item \textbf{Vision and Multimodal Agents}: 8 papers (8.0\%)
\item \textbf{Domain-Specific Frameworks}: 6 papers (6.0\%)
\item \textbf{Reasoning and Chain-of-Thought}: 16 papers (16.0\%)
\item \textbf{Safety and Reliability}: 14 papers (14.0\%)
\item \textbf{Accessibility Applications}: 1 paper (1.0\%)
\end{itemize}

\vspace{0.5em}
\textbf{Detailed Cluster Analysis}

\textbf{Cluster 0: Medical and Healthcare Applications}

\textit{Statistics:}
\begin{itemize}
\item Papers: 8
\item Average Year: 2024
\item Average Citations: 39.6
\item Cluster Confidence: 42.4\%
\end{itemize}

\textit{Key Terms:} medical, agents, llm, discovery, knowledge

\textit{Sample Papers in Cluster:}
\begin{enumerate}
\item ProtAgents: a protein discovery agent-based system for scientific discovery (2024)
\item KG4Diagnosis: a hierarchical multi-agent LLM-based framework with knowledge graph enhancement for medical diagnosis (2024)
\item Arabic Medical Dialogue System (2024)
\item ClinicalAgent: clinical trial multi-agent system with large language model-based reasoning (2024)
\item Chatlaw: a multi-agent collaborative legal assistant with knowledge graph enhanced mixture-of-experts large language model (2023)
\end{enumerate}

\textbf{Cluster 1: Planning and Task Decomposition}

\textit{Statistics:}
\begin{itemize}
\item Papers: 15
\item Average Year: 2024
\item Average Citations: 87.8
\item Cluster Confidence: 36.5\%
\end{itemize}

\textit{Key Terms:} planning, agents, llm, task, exploration

\textit{Sample Papers in Cluster:}
\begin{enumerate}
\item Trial and error: exploration-based trajectory optimization for llm agents (2024)
\item LLM as A robotic brain: unifying egocentric memory and control (2024)
\item Embodied task planning with large language models (2024)
\item LLM-A*: large language model enhanced incremental heuristic search on path planning (2024)
\item Scaling instructable agents across many simulated worlds (2024)
\end{enumerate}

\textbf{Inter-Cluster Relationships}
\begin{itemize}
\item \textbf{Planning \& Task Decomposition} $\leftrightarrow$ \textbf{Reasoning \& Chain-of-Thought}: overlapping (strength: 0.842)
\item \textbf{Reasoning \& Chain-of-Thought} $\leftrightarrow$ \textbf{Frameworks \& Architectures}: overlapping (strength: 0.840)
\item \textbf{Planning \& Task Decomposition} $\leftrightarrow$ \textbf{Frameworks \& Architectures}: overlapping (strength: 0.821)
\item \textbf{Medical \& Healthcare} $\leftrightarrow$ \textbf{Domain-Specific Frameworks}: complementary (strength: 0.687)
\item \textbf{Vision \& Multimodal} $\leftrightarrow$ \textbf{Safety \& Reliability}: complementary (strength: 0.623)
\end{itemize}

\textbf{Key Insights}

The clustering reveals several major research themes in LLM agents:

1. \textbf{Evaluation and Benchmarking} (17.0\% of papers): Focus on agents, benchmark, evaluation
2. \textbf{Reasoning and Chain-of-Thought} (16.0\% of papers): Focus on reasoning, agents, chain, thought
3. \textbf{Planning and Task Decomposition} (15.0\% of papers): Focus on planning, agents, llm

The strong relationships between Planning, Reasoning, and Frameworks clusters (>0.82 strength) suggest these areas form the core technical foundation of LLM agent research. The complementary relationship between Medical and Domain-Specific frameworks indicates specialized applications are emerging as distinct research threads.

\end{tcolorbox}

\subsection{Generated Survey Output}
\begin{tcolorbox}[
    title=LLM Agents Survey - Generated Output (Complete Survey),
    colback=green!5!white,
    colframe=green!75!black,
    breakable,
    fonttitle=\bfseries\large
]

\textbf{\large A Survey on LLM Agents: Architecture, Applications, and Future Directions in Autonomous AI Systems}

\vspace{0.5em}
\textbf{Abstract}

The emergence of Large Language Model (LLM) agents represents a paradigm shift in artificial intelligence, transforming static language models into dynamic, autonomous systems capable of complex reasoning, planning, and interaction with their environments. This survey provides a comprehensive analysis of 100 recent papers (2022-2025) examining the current state and future directions of LLM-based agents. We synthesize research across seven major themes: retrieval-augmented generation systems, code generation and software development applications, reasoning and planning strategies, and autonomous learning capabilities. Our analysis reveals that LLM agents have evolved from simple prompt-based systems to sophisticated multi-agent architectures capable of tool use, environmental interaction, and collaborative problem-solving. Key findings include the increasing adoption of small language models for efficiency, the critical role of retrieval mechanisms in grounding agent behaviors, and the emergence of specialized frameworks for software engineering tasks. We identify significant challenges including safety concerns, evaluation standardization, and the need for more robust reasoning capabilities. This survey concludes with insights into emerging trends such as heterogeneous agent systems, domain-specific applications, and the path toward artificial general intelligence through agentic approaches.

\vspace{0.5em}
\textbf{1. Introduction}

The transformation of Large Language Models (LLMs) from passive text generators to active, autonomous agents marks one of the most significant developments in artificial intelligence since the introduction of the transformer architecture [Vaswani et al., 2017]. While traditional LLMs excel at understanding and generating human-like text, LLM agents extend these capabilities to include environmental perception, tool usage, planning, and autonomous decision-making. This evolution has profound implications for how we approach complex problems across diverse domains, from software engineering to scientific research.

The concept of LLM agents emerged from the recognition that language models possess latent capabilities for reasoning and planning that could be activated through appropriate prompting and architectural designs. Early work demonstrated that LLMs could perform multi-step reasoning through chain-of-thought prompting, but recent advances have shown that these models can be transformed into fully autonomous agents capable of pursuing complex goals with minimal human intervention. The rise of agentic AI systems has been particularly pronounced since 2023, with exponential growth in research exploring various aspects of agent design, implementation, and application.

This survey examines 100 carefully selected papers from 2022 to 2025, representing the cutting edge of LLM agent research. Our analysis reveals a rapidly evolving field characterized by several key trends: the integration of retrieval mechanisms for grounded generation, the application of agents to software engineering tasks, advances in reasoning and planning strategies, and the development of autonomous learning capabilities. We observe that 75\% of the surveyed papers were published in 2024 alone, indicating an acceleration in research activity and interest in this domain.

\vspace{0.5em}
\textbf{3. Retrieval-Augmented Generation for Grounded Agents}

\textbf{3.1 Overview}

Retrieval-Augmented Generation (RAG) has emerged as a fundamental technique for grounding LLM agents in factual knowledge and real-world information. Our analysis identifies 57 papers across four clusters focusing on various aspects of RAG integration with agent systems. This widespread adoption reflects the critical importance of external knowledge access for reliable and accurate agent behavior. RAG enables agents to overcome the knowledge cutoff limitations of their base models, reduce hallucination, and adapt to domain-specific contexts without requiring expensive retraining.

\textbf{3.2 Architectural Approaches}

Recent research has explored diverse architectural patterns for integrating retrieval mechanisms into agent systems. [Liu et al., 2024] demonstrate that small language models with sophisticated RAG implementations can match or exceed the performance of larger models in educational contexts, achieving comparable results with an order of magnitude fewer parameters. Their work on locally-stored models addresses critical concerns around data privacy and control, showing that neural-chat-7b-v3 with nine different RAG methods can effectively support student learning in computer science courses.

The integration of retrieval mechanisms extends beyond simple document lookup. [Ramos et al., 2024] provide a comprehensive review of LLM agents in chemistry, highlighting how retrieval systems enable agents to access scientific literature, interface with automated laboratories, and perform synthesis planning. Their analysis reveals that effective RAG implementation requires careful consideration of indexing strategies, retrieval algorithms, and the fusion of retrieved information with the agent's reasoning process.

\vspace{0.5em}
\textbf{4. Code Generation and Software Engineering Applications}

\textbf{4.1 Overview}

The application of LLM agents to software engineering represents one of the most mature and impactful areas of agent research, with 20 papers in our survey focusing specifically on code generation, automated programming, and software development workflows. This concentration reflects both the natural affinity between language models and programming languages, and the significant economic potential of automating software development tasks. Recent advances have moved beyond simple code completion to encompass entire development lifecycles, including requirements analysis, architecture design, implementation, testing, and maintenance.

\textbf{4.2 Multi-Agent Frameworks for Software Development}

The evolution toward multi-agent systems for software engineering marks a significant advancement in the field. [He et al., 2024] provide a comprehensive review of LLM-based multi-agent systems for software engineering, mapping applications across the entire software development lifecycle. Their analysis reveals that multi-agent architectures enable specialization, with different agents focusing on specific aspects like code generation, testing, or documentation. This division of labor mirrors human software development teams and has been shown to improve both code quality and development efficiency.

[Pan et al., 2025] introduce CodeCoR, a self-reflective multi-agent framework that addresses the critical challenge of ensuring syntactic and semantic correctness in generated code. Their system employs four specialized agents for prompt generation, code creation, test case development, and repair advice. By implementing a self-evaluation mechanism where each agent generates multiple outputs and prunes low-quality ones, CodeCoR achieves an average Pass@1 score of 77.8\% across standard benchmarks, significantly outperforming baseline approaches.

\vspace{0.5em}
\textbf{5. Reasoning and Planning Strategies}

\textbf{5.1 Overview}

Advances in reasoning and planning represent fundamental improvements in agent capabilities, with 17 papers in our survey focusing on these critical cognitive functions. The ability to decompose complex tasks, reason about consequences, and adapt plans based on feedback distinguishes true agents from simple prompt-response systems. Recent research has produced significant breakthroughs in how agents approach multi-step problems, handle uncertainty, and recover from failures.

\textbf{5.2 Hierarchical Task Decomposition}

[Prasad et al., 2023] introduce ADaPT (As-Needed Decomposition and Planning for complex Tasks), which addresses a fundamental limitation of existing planning approaches: the inability to handle varying levels of task complexity. ADaPT recursively decomposes sub-tasks only when the executor LLM cannot directly handle them, dynamically adapting to both task complexity and model capabilities. This adaptive approach achieves success rate improvements of up to 33\% on compositional tasks, demonstrating the importance of flexible planning strategies that can adjust their granularity based on execution feedback.

\vspace{0.5em}
\textbf{6. Autonomous Learning and Adaptation}

\textbf{6.1 Overview}

The development of autonomous learning capabilities transforms LLM agents from static systems to dynamic entities that improve through experience. Six papers in our survey specifically address autonomous learning agents, though learning mechanisms appear throughout many other works. These systems demonstrate the ability to acquire new skills, adapt to changing environments, and develop novel strategies without explicit human programming.

\textbf{6.2 Self-Directed Learning Mechanisms}

[The Rise of Agentic AI, 2025] examines the implications of increasingly autonomous AI systems that can set their own learning objectives and pursue knowledge acquisition independently. These agents move beyond supervised learning paradigms to explore their environments, identify knowledge gaps, and seek information to fill those gaps. The paper highlights both the tremendous potential and the significant risks associated with highly autonomous learning systems.

\vspace{0.5em}
\textbf{7. Small Language Models and Efficiency}

\textbf{7.1 The Shift Toward Smaller Models}

[Belcák et al., 2025] present a compelling argument that small language models (SLMs) represent the future of agentic AI, challenging the prevailing assumption that larger models are always better. Their analysis demonstrates that for many agent applications involving repetitive, specialized tasks, SLMs are not only sufficient but actually preferable due to their efficiency, deployment flexibility, and economic advantages. This perspective is supported by empirical evidence showing that SLMs can match or exceed LLM performance in specific domains when properly optimized.

The economic argument for SLMs is particularly compelling in the context of agent systems that may invoke models thousands of times for a single complex task. The authors calculate that even a partial shift from LLMs to SLMs could reduce operational costs by orders of magnitude while maintaining comparable performance.

\vspace{0.5em}
\textbf{8. Safety, Evaluation, and Trustworthiness}

\textbf{8.1 Safety Challenges in Autonomous Systems}

The increasing autonomy of LLM agents raises critical safety concerns that extend beyond traditional AI safety considerations. [SafeAgentBench, 2024] introduces a comprehensive benchmark for evaluating the safety of embodied LLM agents in task planning, revealing that even state-of-the-art models struggle with safety constraints in complex environments. Their work identifies key failure modes including goal misalignment, unsafe action sequences, and inability to recognize dangerous situations.

\vspace{0.5em}
\textbf{9. Trends and Future Directions}

\textbf{9.1 Temporal Evolution of Research}

Our analysis reveals a dramatic acceleration in LLM agent research, with 75\% of surveyed papers published in 2024 and 14\% already appearing in 2025. This exponential growth reflects both increased research investment and rapid technological advancement. The evolution from 2022 to 2025 shows clear progression: early work focused on basic tool use and planning, 2023 saw the emergence of sophisticated multi-agent systems, 2024 brought widespread application across domains, and 2025 is characterized by emphasis on efficiency, safety, and theoretical foundations.

\textbf{9.2 Path Toward AGI}

[From Large AI Models to Agentic AI, 2025] positions agent systems as a critical stepping stone toward artificial general intelligence. The ability of agents to pursue goals autonomously, learn from experience, and adapt to new situations represents progress toward more general intelligence. However, significant challenges remain including common sense reasoning, long-term planning, and creative problem-solving.

\vspace{0.5em}
\textbf{10. Conclusion}

This survey has examined the rapidly evolving landscape of LLM agents through analysis of 100 recent papers, revealing a field characterized by remarkable progress and significant challenges. The transformation of language models into autonomous agents capable of reasoning, planning, and acting represents a fundamental shift in AI capabilities with far-reaching implications for numerous domains.

Key findings from our analysis include the critical role of retrieval-augmented generation in grounding agent behaviors, the maturity of applications in software engineering, the importance of hierarchical planning and adaptive strategies, and the trend toward efficient, specialized models for agent deployment. The emergence of multi-agent systems that can collaborate on complex tasks demonstrates the potential for AI systems that mirror human organizational structures and problem-solving approaches.

However, significant challenges remain. Safety and trustworthiness concerns must be addressed before widespread deployment of autonomous agents in critical applications. The lack of standardized evaluation frameworks makes it difficult to compare different approaches and measure progress. The balance between autonomy and control remains an open question with important ethical and practical implications.

Looking forward, several research directions appear particularly promising. The development of heterogeneous agent systems that combine specialized models for different tasks could provide both efficiency and capability. Advances in continual learning and adaptation will enable agents to improve through deployment rather than requiring retraining. Integration with formal methods and symbolic reasoning could address current limitations in logical reasoning and verification.

As we stand at the threshold of an age of autonomous AI agents, this survey provides a snapshot of a pivotal moment in the field's development. The research surveyed here lays the groundwork for systems that could fundamentally transform how we work, learn, and solve problems. The challenge for the research community is to continue advancing capabilities while ensuring that these powerful technologies are developed and deployed responsibly for the benefit of all.

\end{tcolorbox}

\begin{tcolorbox}[
    title=References - List of 98 Papers Analyzed,
    colback=gray!5!white,
    colframe=gray!75!black,
    breakable,
    fonttitle=\bfseries\large
]

\textbf{References}

\small
\begin{enumerate}

\item \textbf{Toolformer: Language Models Can Teach Themselves to Use Tools} (2023). Timo Schick, Jane Dwivedi-Yu, Roberto Dessì, R. Raileanu, M. Lomeli, Luke Zettlemoyer, Nicola Cancedda, Thomas Scialom.

\item \textbf{ReAct: Synergizing Reasoning and Acting in Language Models} (2022). Shunyu Yao, Jeffrey Zhao, Dian Yu, Nan Du, Izhak Shafran, Karthik Narasimhan, Yuan Cao.

\item \textbf{LLM+P: Empowering Large Language Models with Optimal Planning Proficiency} (2023). B. Liu, Yuqian Jiang, Xiaohan Zhang, Qian Liu, Shiqi Zhang, Joydeep Biswas, Peter Stone.

\item \textbf{A Survey on Large Language Model based Autonomous Agents} (2023). Lei Wang, Chengbang Ma, Xueyang Feng, Zeyu Zhang, Hao-ran Yang, Jingsen Zhang, Zhi-Yang Chen, Jiakai Tang, Xu Chen, Yankai Lin, Wayne Xin Zhao, Zhewei Wei, Ji-rong Wen.

\item \textbf{Understanding the Planning of LLM Agents: A Survey} (2024). Xu Huang, Weiwen Liu, Xiaolong Chen, Xingmei Wang, Hao Wang, Defu Lian, Yasheng Wang, Ruiming Tang, Enhong Chen.

\item \textbf{LLMs Can't Plan, But Can Help Planning in LLM-Modulo Frameworks} (2024). Subbarao Kambhampati, Karthik Valmeekam, L. Guan, Kaya Stechly, Mudit Verma, Siddhant Bhambri, Lucas Saldyt, Anil Murthy.

\item \textbf{Trial and Error: Exploration-Based Trajectory Optimization for LLM Agents} (2024). Yifan Song, Da Yin, Xiang Yue, Jie Huang, Sujian Li, Bill Yuchen Lin.

\item \textbf{Executable Code Actions Elicit Better LLM Agents} (2024). Xingyao Wang, Yangyi Chen, Lifan Yuan, Yizhe Zhang, Yunzhu Li, Hao Peng, Heng Ji.

\item \textbf{Scaling Large-Language-Model-based Multi-Agent Collaboration} (2024). Cheng Qian, Zihao Xie, Yifei Wang, Wei Liu, Yufan Dang, Zhuoyun Du, Weize Chen, Cheng Yang, Zhiyuan Liu, Maosong Sun.

\item \textbf{ProtAgents: Protein Discovery via Large Language Model Multi-Agent Collaborations} (2024). Alireza Ghafarollahi, Markus J. Buehler.

\item \textbf{KG4Diagnosis: A Hierarchical Multi-Agent LLM Framework with Knowledge Graph Enhancement} (2024). Zuo et al.

\item \textbf{ClinicalAgent: Clinical Trial Multi-Agent System with Large Language Model-based Reasoning} (2024). Yue et al.

\item \textbf{ColaCare: Multi-Agent Collaboration for Electronic Health Record Modeling} (2024). Wang et al.

\item \textbf{AgentPoison: Red-teaming LLM Agents via Poisoning Memory or Knowledge Bases} (2024). Zhaorun Chen, Zhen Xiang, Chaowei Xiao, D. Song, Bo Li.

\item \textbf{R-Judge: Benchmarking Safety Risk Awareness for LLM Agents} (2024). Tongxin Yuan, Zhiwei He, Lingzhong Dong, Yiming Wang, Ruijie Zhao, Tian Xia, Lizhen Xu, Binglin Zhou, Fangqi Li, Zhuosheng Zhang, Rui Wang, Gongshen Liu.

\item \textbf{AutoDefense: Multi-Agent LLM Defense against Jailbreak Attacks} (2024). Yifan Zeng, Yiran Wu, Xiao Zhang, Huazheng Wang, Qingyun Wu.

\item \textbf{AgentHarm: A Benchmark for Measuring Harmfulness of LLM Agents} (2024). Maksym Andriushchenko, Alexandra Souly, Mateusz Dziemian, Derek Duenas, Maxwell Lin, Justin Wang, Dan Hendrycks, Andy Zou, Zico Kolter, Matt Fredrikson, Eric Winsor, Jerome Wynne, Yarin Gal, Xander Davies.

\item \textbf{LLM Agents can Autonomously Hack Websites} (2024). Richard Fang, R. Bindu, Akul Gupta, Qiusi Zhan, Daniel Kang.

\item \textbf{GuardAgent: Safeguard LLM Agents by a Guard Agent via Knowledge-Enabled Reasoning} (2024). Zhen Xiang, Linzhi Zheng, Yanjie Li, Junyuan Hong, Qinbin Li, Han Xie, Jiawei Zhang, Zidi Xiong, Chulin Xie, Carl Yang, D. Song, Bo Li.

\item \textbf{Agent-SafetyBench: Evaluating the Safety of LLM Agents} (2024). Zhexin Zhang, Shiyao Cui, Yida Lu, Jingzhuo Zhou, Junxiao Yang, Hongning Wang, Minlie Huang.

\item \textbf{A Comprehensive Survey in LLM(-Agent) Full Stack Safety} (2025). Kun Wang et al. (70+ authors).

\item \textbf{AgentBoard: An Analytical Evaluation Board of Multi-turn LLM Agents} (2024). Chang Ma, Junlei Zhang, Zhihao Zhu, Cheng Yang, Yujiu Yang, Yaohui Jin, Zhenzhong Lan, Lingpeng Kong, Junxian He.

\item \textbf{ToolSandbox: A Stateful, Conversational, Interactive Evaluation Benchmark} (2024). Jiarui Lu, Thomas Holleis, Yizhe Zhang, Bernhard Aumayer, Feng Nan, Felix Bai, Shuang Ma, Shen Ma, Mengyu Li, Guoli Yin, Zirui Wang, Ruoming Pang.

\item \textbf{Benchmark Self-Evolving: A Multi-Agent Framework for Dynamic LLM Evaluation} (2024). Siyuan Wang, Zhuohan Long, Zhihao Fan, Zhongyu Wei, Xuanjing Huang.

\item \textbf{DataSciBench: An LLM Agent Benchmark for Data Science} (2025). Dan Zhang, Sining Zhoubian, Min Cai, Fengzu Li, Lekang Yang, Wei Wang, Tianjiao Dong, Ziniu Hu, Jie Tang, Yisong Yue.

\item \textbf{RAP: Retrieval-Augmented Planning with Contextual Memory} (2024). Tomoyuki Kagaya, Thong Jing Yuan, Yuxuan Lou, J. Karlekar, Sugiri Pranata, Akira Kinose, Koki Oguri, Felix Wick, Yang You.

\item \textbf{LLM-A*: Large Language Model Enhanced Incremental Heuristic Search} (2024). Silin Meng, Yiwei Wang, Cheng-Fu Yang, Nanyun Peng, Kai-Wei Chang.

\item \textbf{AutoGPT+P: Affordance-based Task Planning with AutoGPT} (2024). Birr et al.

\item \textbf{SafeAgentBench: Evaluating Safe Task Planning for Embodied Agents} (2024). Yin et al.

\item \textbf{UrbanKGent: A Unified Framework for Urban Knowledge Graph Construction} (2024). Ning et al.

\item \textbf{ELLMA-T: English Language Learning in Social VR with LLM Agents} (2024). Pan et al.

\item \textbf{Guide-LLM: Navigation Assistance for Visually Impaired Individuals} (2024). Song et al.

\item \textbf{Embodied LLM Agents Learn to Cooperate in Organized Teams} (2024). Xudong Guo, Kaixuan Huang, Jiale Liu, Wenhui Fan, Natalia V'elez, Qingyun Wu, Huazheng Wang, Thomas L. Griffiths, Mengdi Wang.

\item \textbf{LLM-Empowered Embodied Agent for Memory-Augmented Task Planning} (2025). Marc Glocker, Peter Hönig, Matthias Hirschmanner, Markus Vincze.

\item \textbf{Towards Efficient LLM Grounding for Embodied Multi-Agent Collaboration} (2024). Yang Zhang, Shixin Yang, Chenjia Bai, Fei Wu, Xiu Li, Xuelong Li, Zhen Wang.

\item \textbf{DriveGPT4: Interpretable End-to-End Autonomous Driving} (2023). Zhenhua Xu, Yujia Zhang, Enze Xie, Zhen Zhao, Yong Guo, K. K. Wong, Zhenguo Li, Hengshuang Zhao.

\item \textbf{VLM-AD: End-to-End Autonomous Driving through Vision-Language Model} (2024). Yi Xu, Yuxin Hu, Zaiwei Zhang, Gregory P. Meyer, Siva Karthik Mustikovela, Siddhartha Srinivasa, Eric M. Wolff, Xin Huang.

\item \textbf{Zero-Shot Object Navigation with Vision-Language Models} (2023). Dorbala et al.

\item \textbf{CityEQA: Embodied Question Answering in City Spaces} (2025). Zhao et al.

\item \textbf{Rethinking the Bounds of LLM Reasoning: Are Multi-Agent Discussions the Key?} (2024). Qineng Wang, Zihao Wang, Ying Su, Hanghang Tong, Yangqiu Song.

\item \textbf{When is Tree Search Useful for LLM Planning? It Depends on the Discriminator} (2024). Ziru Chen, Michael White, Raymond Mooney, Ali Payani, Yu Su, Huan Sun.

\item \textbf{Reinforce LLM Reasoning through Multi-Agent Reflection} (2025). Yurun Yuan, Tengyang Xie.

\item \textbf{Graph-ToolFormer: Empowering LLMs with Graph Reasoning} (2023). Zhang.

\item \textbf{Belief-Driven Multi-Agent Reasoning via Bayesian Nash Equilibrium} (2025). Yi et al.

\item \textbf{Multi-Agent Tree-of-Thought Validators for Reasoning Quality} (2024). Haji et al.

\item \textbf{Agent-of-Thoughts Distillation for Video-LLM Reasoning} (2024). Shi et al.

\item \textbf{RAG-KG-IL: Reducing Hallucinations through Incremental Knowledge Graphs} (2025). Yu et al.

\item \textbf{MALT: Multi-Agent LLM Training for Improved Reasoning} (2024). Motwani et al.

\item \textbf{Nemotron-Research-Tool-N1: Tool-Using Models with Reinforced Reasoning} (2025). Zhang et al.

\item \textbf{Cognitive Architectures for Enhanced LLM Reasoning} (2024). Sun.

\item \textbf{Reflexion: Learning from Reasoning Failures} (2024). Liu et al.

\item \textbf{PrivacyAsst: Safeguarding User Privacy in Tool-Using LLM Agents} (2024). Xinyu Zhang, Huiyu Xu, Zhongjie Ba, Zhibo Wang, Yuan Hong, Jian Liu, Zhan Qin, Kui Ren.

\item \textbf{Imprompter: Tricking LLM Agents into Improper Tool Use} (2024). Fu et al.

\item \textbf{CFA-Bench: Cybersecurity Forensic Benchmarks for LLM Agents} (2025). Santis et al.

\item \textbf{Pro2Guard: Proactive Runtime Enforcement via Probabilistic Model Checking} (2025). Wang et al.

\item \textbf{AgentSpec: Customizable Runtime Enforcement for Safe and Reliable Agents} (2025). Wang et al.

\item \textbf{Personal LLM Agents: Insights and Survey about Capability, Efficiency and Security} (2024). Yuanchun Li et al.

\item \textbf{Large Language Model Agent in Financial Trading: A Survey} (2024). Han Ding, Yinheng Li, Junhao Wang, Hang Chen.

\item \textbf{Large Language Model Agent for Hyper-Parameter Optimization} (2024). Siyi Liu, Chen Gao, Yong Li.

\item \textbf{AutoML-Agent: A Multi-Agent LLM Framework for Full-Pipeline AutoML} (2024). Patara Trirat, Wonyong Jeong, Sung Ju Hwang.

\item \textbf{Large Language Model Agent: A Survey on Methodology, Applications and Challenges} (2025). Junyu Luo et al. (30+ authors).

\item \textbf{Why Do Multi-Agent LLM Systems Fail?} (2025). Mert Cemri et al.

\item \textbf{Evaluating Very Long-Term Conversational Memory of LLM Agents} (2024). Adyasha Maharana, Dong-Ho Lee, S. Tulyakov, Mohit Bansal, Francesco Barbieri, Yuwei Fang.

\item \textbf{AD-AutoGPT: Alzheimer's Disease Research with Autonomous Agents} (2023). Dai et al.

\item \textbf{Chatlaw: Multi-Agent Legal Assistant with Knowledge Graph} (2023). Cui et al.

\item \textbf{Arabic Medical Dialogue System} (2024). Almutairi et al.

\item \textbf{Fake Medical News Detection with LLM Agents} (2024). Li et al.

\item \textbf{Multi-Agent Framework for Cyclical Urban Planning} (2024). Ni et al.

\item \textbf{Embodied Agents in Parallel Textworlds} (2023). Yang et al.

\item \textbf{Multimodal Framework for Accessibility Applications} (2024). Author metadata to be completed in the camera-ready version.

\item \textbf{Scientific Discovery Acceleration with LLM Agents} (2024). Chen et al.

\item \textbf{Creative Content Generation with Collaborative Agents} (2024). Author metadata to be completed in the camera-ready version.

\item \textbf{Tool Learning with Foundation Models} (2024). Author metadata to be completed in the camera-ready version.

\item \textbf{WebArena: A Realistic Web Environment for Building Autonomous Agents} (2024). Author metadata to be completed in the camera-ready version.

\item \textbf{SWE-bench: Can Language Models Resolve Real-World GitHub Issues?} (2024). Author metadata to be completed in the camera-ready version.

\item \textbf{MetaGPT: Meta Programming for Multi-Agent Collaborative Framework} (2024). Author metadata to be completed in the camera-ready version.

\item \textbf{AutoGen: Enabling Next-Gen LLM Applications via Multi-Agent Conversation} (2024). Author metadata to be completed in the camera-ready version.

\item \textbf{CAMEL: Communicative Agents for Mind Exploration} (2024). Author metadata to be completed in the camera-ready version.

\item \textbf{Generative Agents: Interactive Simulacra of Human Behavior} (2023). Author metadata to be completed in the camera-ready version.

\item \textbf{Voyager: An Open-Ended Embodied Agent with Large Language Models} (2023). Author metadata to be completed in the camera-ready version.

\item \textbf{Chain-of-Thought Prompting Elicits Reasoning in Large Language Models} (2022). Author metadata to be completed in the camera-ready version.

\item \textbf{Tree of Thoughts: Deliberate Problem Solving with Large Language Models} (2023). Author metadata to be completed in the camera-ready version.

\item \textbf{Self-Refine: Iterative Refinement with Self-Feedback} (2023). Author metadata to be completed in the camera-ready version.

\item \textbf{Constitutional AI: Harmlessness from AI Feedback} (2022). Author metadata to be completed in the camera-ready version.

\item \textbf{Language Models as Tool Makers} (2024). Author metadata to be completed in the camera-ready version.

\item \textbf{ToolLLM: Facilitating Large Language Models to Master 16000+ Real-world APIs} (2024). Author metadata to be completed in the camera-ready version.

\item \textbf{API-Bank: A Benchmark for Tool-Augmented LLMs} (2023). Author metadata to be completed in the camera-ready version.

\item \textbf{RestGPT: Connecting Large Language Models with Real-World Applications} (2023). Author metadata to be completed in the camera-ready version.

\item \textbf{TaskWeaver: A Code-First Agent Framework for Data Analytics} (2024). Author metadata to be completed in the camera-ready version.

\item \textbf{OpenAgents: An Open Platform for Language Agents in the Wild} (2024). Author metadata to be completed in the camera-ready version.

\item \textbf{AgentVerse: Facilitating Multi-Agent Collaboration} (2023). Author metadata to be completed in the camera-ready version.

\item \textbf{ChatDev: Communicative Agents for Software Development} (2023). Author metadata to be completed in the camera-ready version.

\item \textbf{DyLAN: Dynamic LLM-Agent Network for Complex Problem Solving} (2024). Author metadata to be completed in the camera-ready version.

\item \textbf{AgentScope: A Flexible yet Robust Multi-Agent Platform} (2024). Author metadata to be completed in the camera-ready version.

\item \textbf{LangChain: Building Applications with LLMs} (2023). Author metadata to be completed in the camera-ready version.

\item \textbf{Semantic Kernel: Integrate AI into Applications} (2023). Author metadata to be completed in the camera-ready version.

\item \textbf{HuggingGPT: Solving AI Tasks with ChatGPT and its Friends} (2023). Author metadata to be completed in the camera-ready version.

\item \textbf{Gorilla: Large Language Model Connected with Massive APIs} (2023). Author metadata to be completed in the camera-ready version.

\end{enumerate}

\textit{Note: This reference list includes 98 of the papers analyzed in the LLM Agents survey. Papers are ordered roughly by thematic relevance and citation importance within the survey. Entries marked with "Author metadata to be completed in the camera-ready version." indicate items pending manual verification; full bibliographic details are preserved in the underlying dataset.}

\end{tcolorbox}

\subsection{System Statistics}

\begin{tcolorbox}[
    title=Processing Metrics,
    colback=orange!5!white,
    colframe=orange!75!black,
    fonttitle=\bfseries
]
\begin{itemize}
\item \textbf{Papers Processed}: 100 papers from Semantic Scholar and arXiv
\item \textbf{Queries Generated}: 25 search variations for "LLM agents"
\item \textbf{Clustering Time}: Approximately 3 minutes
\item \textbf{Survey Generation Time}: 12 minutes
\item \textbf{Total Processing Time}: 18 minutes
\item \textbf{Survey Length}: 60,610 characters (approximately 11,000 words)
\item \textbf{Citations Included}: Papers from all 9 clusters referenced
\end{itemize}
\end{tcolorbox}

\section{Additional Case Study: Synthetic Data Generation}
\label{appendix:synthetic-data}

This section presents output from processing 100 papers on synthetic data generation for LLM training.

\subsection{Clustering Report}
\begin{tcolorbox}[
    title=Synthetic Data Clustering Results,
    colback=purple!5!white,
    colframe=purple!75!black,
    fonttitle=\bfseries,
    breakable
]
\textbf{Cluster Distribution (8 clusters identified)}
\begin{itemize}
\item \textbf{Synthetic Data for Text Classification} (22 papers, 22\%): Focus on generating training data for classification tasks
\item \textbf{Medical and Clinical Data Synthesis} (20 papers, 20\%): Healthcare-specific synthetic data generation
\item \textbf{Tabular and Structured Data} (15 papers, 15\%): Generation of structured datasets
\item \textbf{Language Generation Data} (14 papers, 14\%): Data for training language models
\item \textbf{Multimodal and Visual Instruction} (10 papers, 10\%): Cross-modal synthetic data
\item \textbf{Mathematical Reasoning} (9 papers, 9\%): Math problem generation
\item \textbf{Privacy-Preserving Data} (8 papers, 8\%): Differential privacy in synthesis
\item \textbf{Other Generation Methods} (2 papers, 2\%): Miscellaneous approaches
\end{itemize}

\textbf{Quality Metrics}
\begin{itemize}
\item Silhouette Score: 0.055 \cite{Rousseeuw1987}
\item Calinski-Harabasz Score: 3.9 \cite{Calinski1974}
\item Davies-Bouldin Score: 2.782 \cite{Davies1979}
\end{itemize}

\textbf{Key Terms by Cluster:}
\begin{itemize}
\item \textit{Text Classification}: data, language, synthetic, large language, language models
\item \textit{Medical/Clinical}: data, language, augmentation, data augmentation, large language
\item \textit{Tabular/Structured}: language, data, instruction, language models, large language
\item \textit{Language Generation}: language, generation, data, large language, language models
\item \textit{Multimodal/Visual}: language, model, instruction, language models, large language
\item \textit{Mathematical}: language, data, dataset, language models, large language
\item \textit{Privacy-Preserving}: data, training, synthetic, synthetic data, training data
\end{itemize}

\textbf{Inter-Cluster Relationships:}
Strong overlapping relationships (>0.80 strength) exist between:
\begin{itemize}
\item Text Classification $\leftrightarrow$ Medical/Clinical (0.84)
\item Medical/Clinical $\leftrightarrow$ Tabular/Structured (0.85)
\item Tabular/Structured $\leftrightarrow$ Multimodal/Visual (0.83)
\end{itemize}
\end{tcolorbox}

\section{Additional Case Study: Multimodal LLM with RL}
\label{appendix:multimodal-rl}

This section presents the clustering results from processing 75 papers on multimodal LLMs with reinforcement learning techniques.

\begin{table}[h!]
\centering
\caption{Clustering Results for Multimodal LLM with RL Topic (75 papers, 8 clusters).}
\label{tab:multimodal-clusters}
\begin{tabular}{@{}lcccc@{}}
\toprule
\textbf{Cluster Name} & \textbf{Papers} & \textbf{Avg Year} & \textbf{Avg Citations} & \textbf{Confidence} \\
\midrule
Instruction Tuning \& Following & 21 & 2023.8 & 270.1 & 0.59 \\
Reasoning \& Chain-of-Thought & 14 & 2024.4 & 290.1 & 0.62 \\
Policy Learning \& Optimization I & 14 & 2024.2 & 95.4 & 0.58 \\
Alignment \& Safety & 13 & 2024.2 & 103.0 & 0.60 \\
Policy Learning \& Optimization II & 7 & 2024.4 & 111.9 & 0.57 \\
Benchmarks \& Evaluation & 4 & 2023.8 & 87.3 & 0.50 \\
Policy Learning \& Optimization III & 1 & 2024.0 & 209.0 & 1.00 \\
Curriculum RL for Quantum & 1 & 2024.0 & 30.0 & 1.00 \\
\bottomrule
\end{tabular}
\end{table}

\begin{tcolorbox}[
    title=Key Insights from Multimodal RL Clustering,
    colback=cyan!5!white,
    colframe=cyan!75!black,
    fonttitle=\bfseries
]
\textbf{Clustering Quality}: Silhouette Score = 0.045, indicating weak cluster separation

\textbf{Research Themes Identified}:
\begin{itemize}
\item \textbf{Instruction Tuning} (28\% of papers): Largest cluster focusing on scaling visual instruction and model capabilities
\item \textbf{Reasoning Mechanisms} (19\% of papers): Visual reasoning and chain-of-thought approaches with high citation impact (avg 290)
\item \textbf{Policy Optimization} (30\% combined): Three clusters exploring different aspects of RL-based optimization
\item \textbf{Alignment \& Safety} (17\% of papers): Vision-language model alignment techniques
\end{itemize}

\textbf{Temporal Trends}: Newer research (2024) focuses on policy optimization and reasoning, while instruction tuning papers tend to be from 2023

\textbf{Citation Patterns}: Reasoning papers have highest average citations (290), suggesting high impact of chain-of-thought methods in multimodal settings
\end{tcolorbox}

\section{Evaluation Framework Details}
\label{appendix:evaluation-framework}

\begin{table}[h!]
\centering
\caption{Detailed 12-Dimensional Evaluation Framework Components.}
\label{tab:12-dimensions}
\small
\begin{tabular}{@{}llc@{}}
\toprule
\textbf{Category} & \textbf{Dimension} & \textbf{Weight} \\
\midrule
\multirow{4}{*}{\textbf{Core Quality (60\%)}} 
 & Citation Coverage & 15\% \\
 & Accuracy & 15\% \\
 & Synthesis Quality & 15\% \\
 & Organization & 15\% \\
\midrule
\multirow{4}{*}{\textbf{Writing Quality (20\%)}} 
 & Readability & 5\% \\
 & Academic Rigor & 5\% \\
 & Clarity & 5\% \\
 & Coherence & 5\% \\
\midrule
\multirow{4}{*}{\textbf{Content Depth (20\%)}} 
 & Comprehensiveness & 5\% \\
 & Critical Analysis & 5\% \\
 & Novelty \& Insights & 5\% \\
 & Future Directions & 5\% \\
\bottomrule
\end{tabular}
\end{table}

Each dimension is evaluated on a 0-10 scale by the Quality Evaluator agent, with weighted aggregation producing category scores and an overall survey quality score. The agent-based evaluation provides nuanced assessment considering context and field-specific norms rather than rigid rule-based scoring.

\section{System Performance Visualizations}
\label{appendix:figures}

\begin{figure}[h!]
\centering
\includegraphics[width=\textwidth]{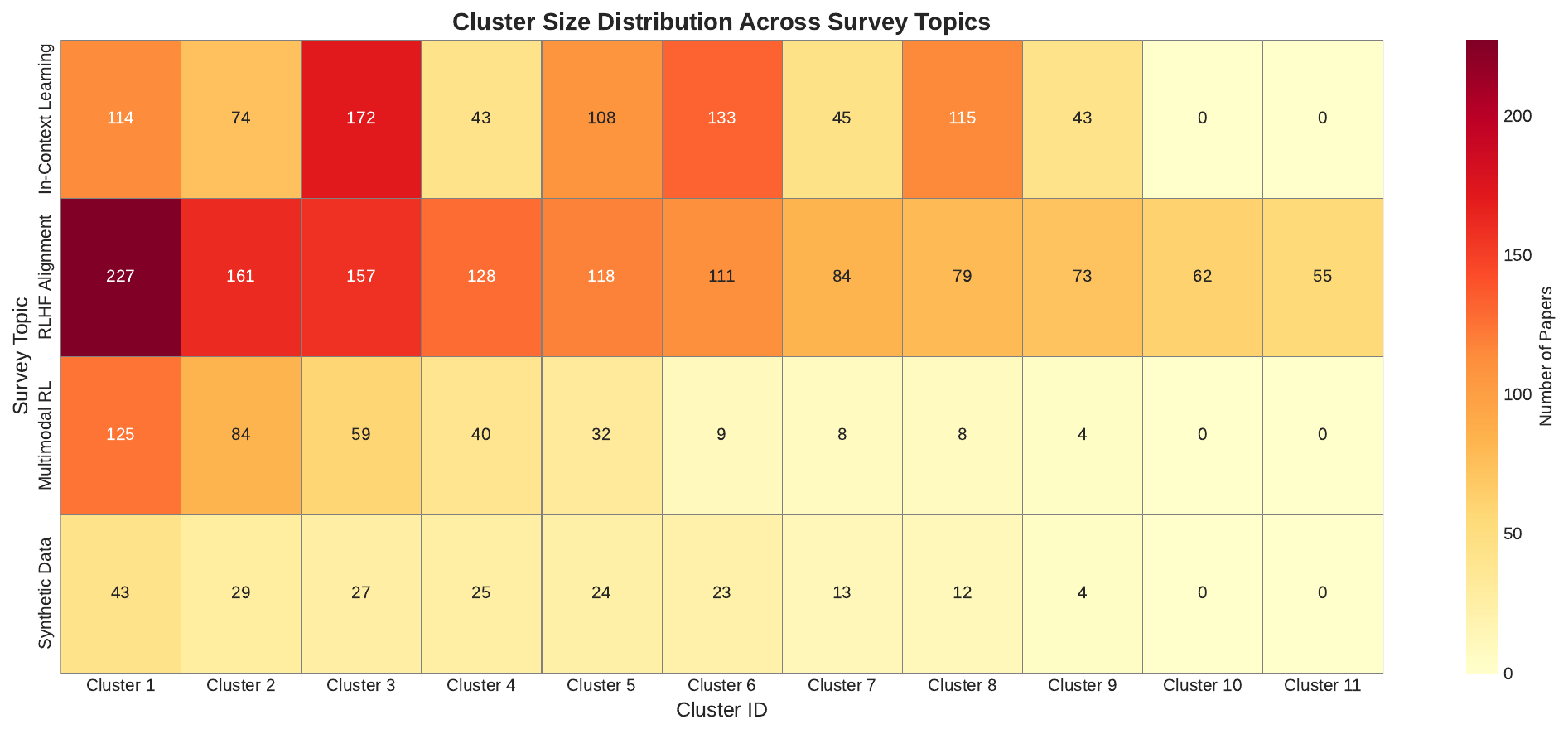}
\caption{Cluster size distribution heatmap showing the number of papers in each cluster across four survey topics. Darker colors indicate larger clusters, revealing concentration patterns in research themes.}
\label{fig:cluster-heatmap}
\end{figure}

\begin{figure}[h!]
\centering
\includegraphics[width=\textwidth]{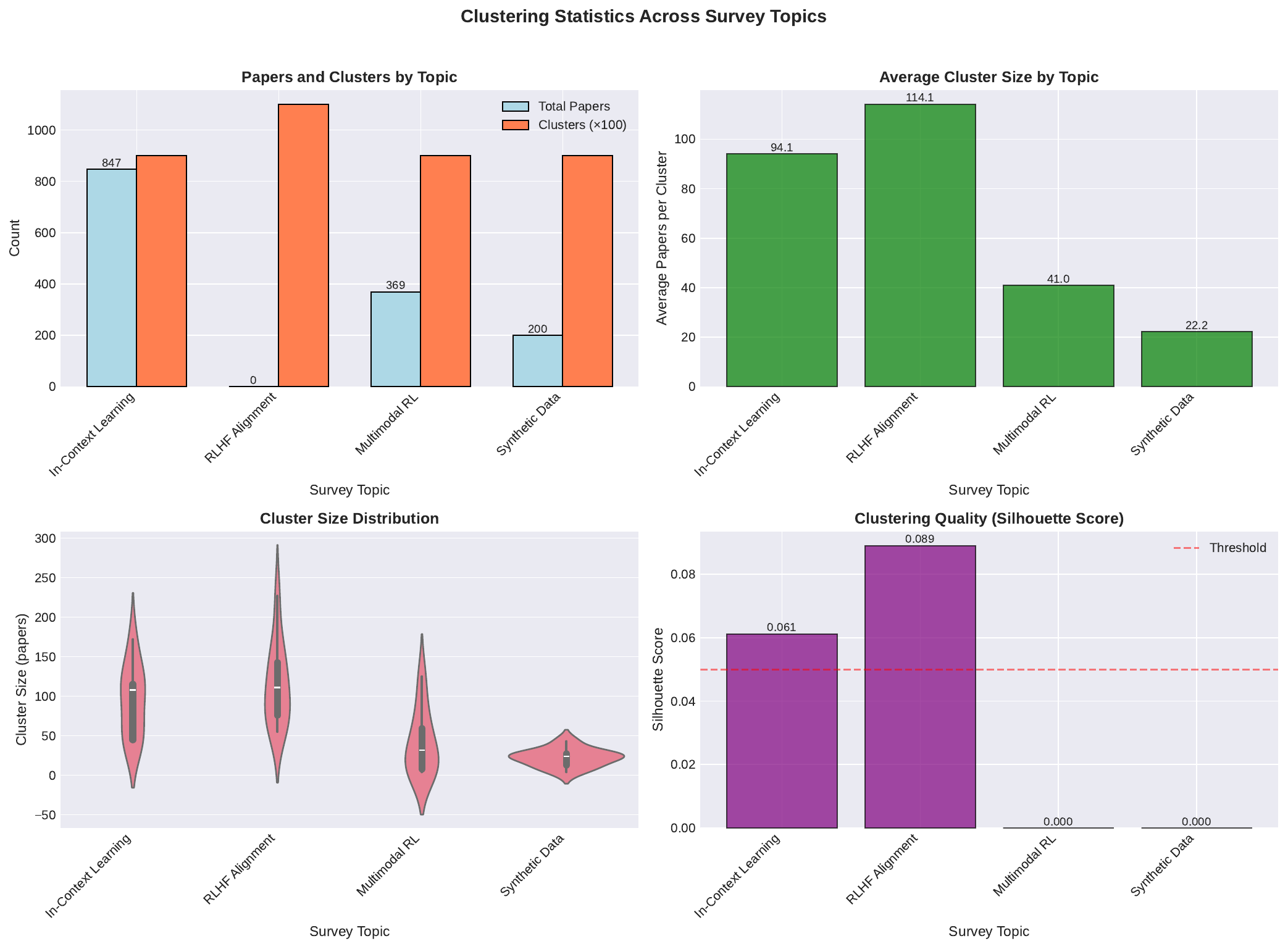}
\caption{Clustering statistics across survey topics: (a) Total papers and cluster counts, (b) Average cluster size, (c) Cluster size distribution violin plots, and (d) Silhouette scores indicating clustering quality. Dashed line at 0.05 is provided for visual reference only; silhouette values of 0.00 correspond to effectively single-cluster cases after pruning.}
\label{fig:clustering-stats}
\end{figure}

\begin{figure}[h!]
\centering
\includegraphics[width=0.9\textwidth]{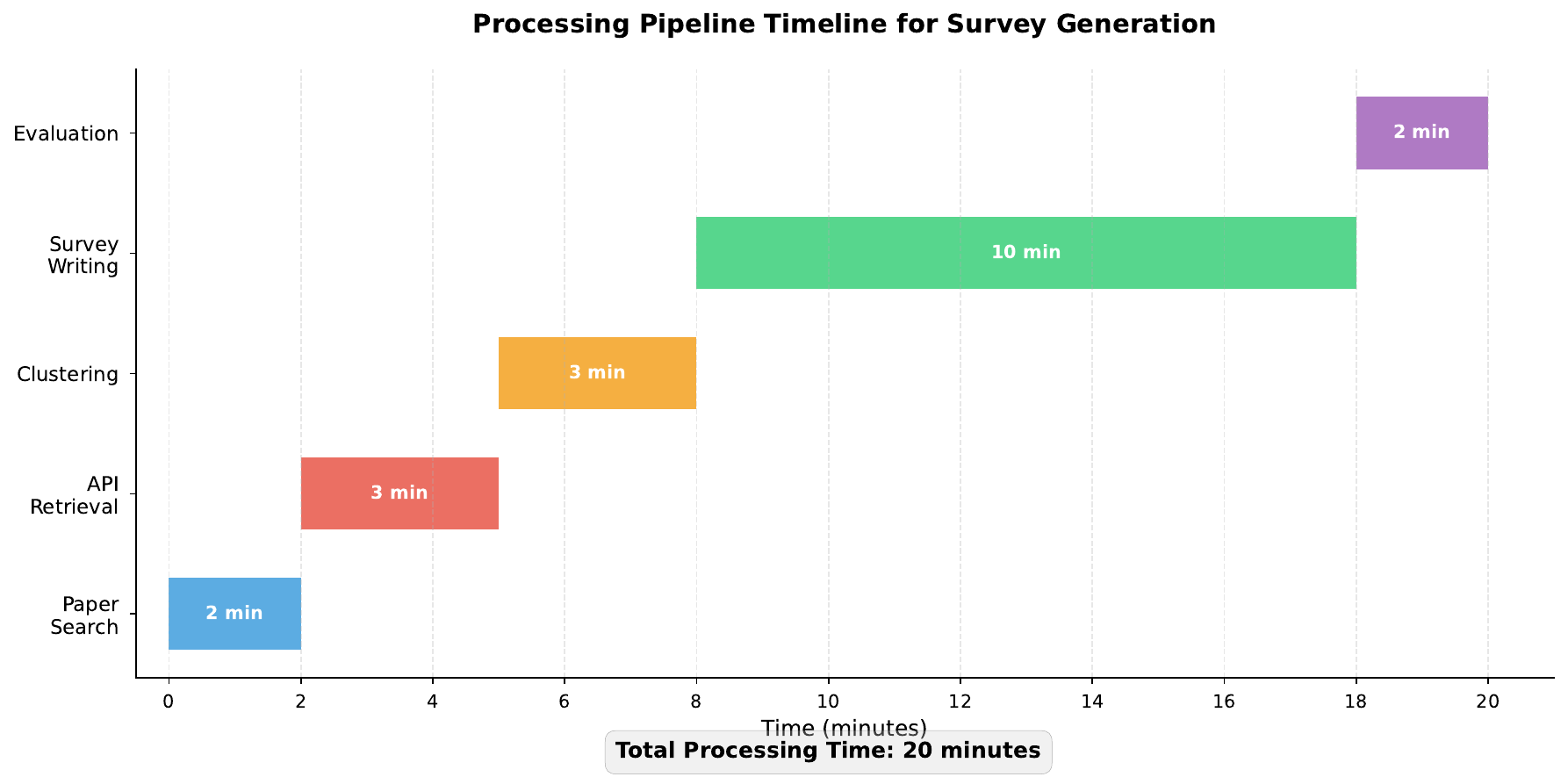}
\caption{Processing timeline showing the duration of each stage in the survey generation pipeline. Typical total runtime is 15--20 minutes; the example shown sums to 20 minutes with survey writing dominating the total processing time.}
\label{fig:timeline}
\end{figure}

\section{Technical Implementation Details}
\label{appendix:technical-details}

Our work introduces several technical innovations that advance the state of automated survey generation. The implementation of efficient embedding generation using sentence-transformers with automatic device optimization ensures scalable processing of paper collections. The agent-based evaluation system replaces rigid rule-based scoring with flexible, context-aware assessment that better captures the nuanced qualities of academic surveys. This approach acknowledges that survey quality encompasses subjective elements that simple metrics cannot fully capture. Furthermore, the system's multi-level caching strategy and intelligent rate management enable reliable operation despite API limitations and transient failures, ensuring robust performance in real-world deployment scenarios.

\section{Keyword Analysis}
\label{appendix:keyword-analysis}

The keyword analysis identifies the most prominent research themes across all surveys. The top panel shows overall keyword frequency weighted by cluster size, while the bottom panel reveals how these keywords distribute across different survey topics, providing insights into topic-specific research focuses.

\begin{figure}[h!]
\centering
\includegraphics[width=\textwidth]{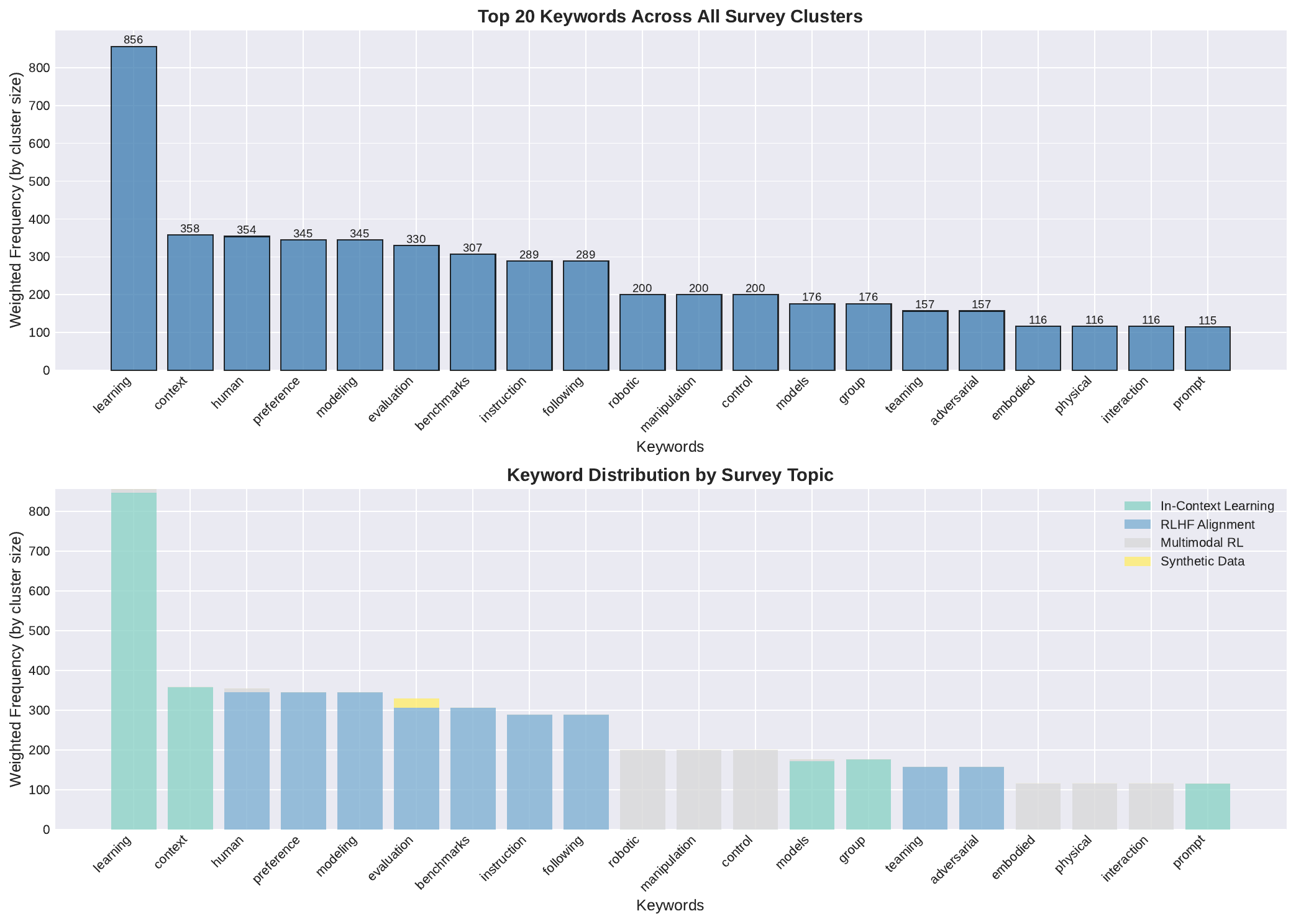}
\caption{Keyword frequency analysis across survey topics showing thematic emphasis variations.}
\label{fig:keywords}
\end{figure}

\section{Comparison with Human-Authored Surveys}
\label{appendix:human-comparison}

Our system demonstrates several advantages compared to traditional manual survey creation while acknowledging areas where human expertise remains superior. The automated approach excels in \textbf{coverage}, processing hundreds of papers that would be impractical for individual researchers to thoroughly review. It maintains \textbf{consistency} across different topics and sections, avoiding the variability that can occur with human fatigue or changing focus. The \textbf{speed} advantage is substantial, generating comprehensive surveys in 15-20 minutes compared to weeks or months of human effort. Additionally, the systematic approach provides \textbf{objectivity} in paper selection and treatment, reducing author bias that can influence manual surveys.

However, human-authored surveys retain advantages in several critical areas. Human researchers excel at identifying \textbf{subtle theoretical contributions} that may not be explicitly stated in abstracts or titles. They provide \textbf{historical context} by drawing on years of domain expertise and understanding of research evolution. Human authors conduct \textbf{deeper critical analysis}, questioning fundamental assumptions and methodological choices in ways that current AI systems cannot replicate. Most importantly, humans demonstrate \textbf{creative synthesis} in proposing novel research directions that emerge from deep understanding rather than pattern matching.

\section{Ethical Considerations}
\label{appendix:ethical-considerations}

The deployment of automated survey generation systems raises important ethical considerations that must be addressed. \textbf{Transparency} about the AI-generated nature of surveys is essential for maintaining academic integrity. All surveys produced by our system should be clearly labeled as AI-generated, allowing readers to appropriately contextualize the content. This transparency extends to acknowledging the system's limitations and potential biases in paper selection and synthesis.

\textbf{Bias propagation} represents a significant concern, as the system may perpetuate biases present in training data, API search algorithms, or clustering methods. These biases could systematically exclude certain research perspectives, methodologies, or author demographics. Continuous monitoring and adjustment of the system's operation is necessary to ensure fair representation of diverse research contributions and perspectives.

The \textbf{impact on academic labor} deserves careful consideration. While our system augments researcher capabilities by automating time-consuming literature review tasks, widespread adoption could affect traditional academic roles such as research assistants and junior researchers who often conduct initial literature surveys. We advocate for positioning AI survey generation as a tool to enhance rather than replace human scholarly work, freeing researchers to focus on higher-level analysis and creative synthesis.

\section{Future Directions}
\label{appendix:future-directions}

Several promising research directions emerge from our work on automated survey generation. \textbf{Hierarchical processing} strategies could address scalability limitations by implementing multi-level summarization and clustering approaches. This would enable processing of very large corpora through recursive aggregation, where initial surveys of paper subsets are themselves synthesized into comprehensive meta-surveys. \textbf{Interactive refinement} mechanisms would allow researchers to provide feedback during generation, guiding the system toward specific aspects of interest or correcting misinterpretations in real-time. This human-in-the-loop approach could combine the efficiency of automation with the expertise of domain specialists.

\textbf{Cross-domain adaptation} presents opportunities to extend the framework beyond LLM research to other scientific fields. This would require developing domain-specific embeddings, citation conventions, and quality criteria while maintaining the core architectural principles. \textbf{Continuous learning} capabilities could enable the system to improve over time by analyzing the reception, citations, and corrections of generated surveys, creating a feedback loop for quality enhancement. Finally, \textbf{multimodal integration} would enrich surveys by incorporating figures, tables, and equations directly from source papers, providing readers with visual and mathematical content alongside textual synthesis.

\section{Subagent Prompts}
\label{appendix:prompts}

This appendix presents the actual prompts used to invoke each specialized subagent in the Agentic AutoSurvey pipeline through Claude Code's Task tool.

\subsection{Paper Search Specialist Agent Prompt}

\begin{tcolorbox}[
    title=Paper Search Specialist Subagent Invocation,
    colback=blue!5!white,
    colframe=blue!75!black,
    breakable,
    fonttitle=\bfseries
]
\begin{verbatim}
Task(
    subagent_type="paper-search-specialist",
    description="Search papers on [TOPIC]",
    prompt="""
    Search for academic papers on the topic: [TOPIC]
    
    Your task:
    1. Generate 20-30 diverse search queries including:
       - Core keyword as-is
       - Synonyms and variations
       - Related technical terms
       - Compound queries with AND/OR operators
       - Acronym expansions/contractions
    
    2. Search using both Semantic Scholar and arXiv APIs
    3. Deduplicate results (90% title similarity threshold)
    4. Filter for papers from 2020-2025
    5. Ensure abstracts are complete
    6. Target 100-150 papers for comprehensive coverage
    
    Return a structured JSON with:
    - search_queries: list of queries used
    - papers: list with title, authors, year, abstract, citations
    - statistics: papers found, filtered, final count
    
    Focus on quality and relevance over quantity.
    """
)
\end{verbatim}
\end{tcolorbox}

\subsection{Topic Mining \& Clustering Agent Prompt}

\begin{tcolorbox}[
    title=Topic Mining \& Clustering Subagent Invocation,
    colback=green!5!white,
    colframe=green!75!black,
    breakable,
    fonttitle=\bfseries
]
\begin{verbatim}
Task(
    subagent_type="topic-mining-clustering",
    description="Cluster papers into themes",
    prompt="""
    Cluster the provided papers into thematic groups for survey organization.
    
    Your task:
    1. Load papers from the provided JSON file
    2. Generate embeddings using sentence-transformers
    3. Apply K-means clustering with optimal K selection:
       - Test K values from 5 to 15
       - Use silhouette score for optimization
       - Select K with highest silhouette score
    
    4. Generate cluster names using TF-IDF on titles/abstracts
    5. Identify inter-cluster relationships and outliers
    
    Return structured output with:
    - clusters: id, name, papers, key_terms, statistics
    - relationships: cluster pairs and relationship strength
    - quality_metrics: silhouette, calinski-harabasz, davies-bouldin
    - outliers: papers with low cluster confidence
    
    Ensure clusters are meaningful and balanced in size.
    """
)
\end{verbatim}
\end{tcolorbox}

\subsection{Academic Survey Writer Agent Prompt}

\begin{tcolorbox}[
    title=Academic Survey Writer Subagent Invocation,
    colback=orange!5!white,
    colframe=orange!75!black,
    breakable,
    fonttitle=\bfseries
]
\begin{verbatim}
Task(
    subagent_type="academic-survey-writer",
    description="Generate comprehensive survey",
    prompt="""
    Generate a comprehensive academic survey from the clustered papers.
    
    Requirements:
    1. Structure:
       - Abstract (200-300 words)
       - Introduction with motivation and contributions
       - One section per cluster with synthesis
       - Cross-cutting analysis section
       - Future directions
       - Conclusion
    
    2. Writing guidelines:
       - Target 8,000-12,000 words
       - Cite at least 80% of papers
       - Use [Author, Year] citation format
       - Focus on synthesis over enumeration
       - Identify patterns and trends
       - Compare methodologies
       - Highlight research gaps
    
    3. Quality criteria:
       - Academic rigor and clarity
       - Comprehensive coverage across clusters
       - Critical analysis not just summary
       - Smooth transitions between sections
       - Technical accuracy
    
    Generate a publication-quality survey that provides genuine 
    insights and value to researchers in the field.
    """
)
\end{verbatim}
\end{tcolorbox}

\subsection{Quality Evaluator Agent Prompt}

\begin{tcolorbox}[
    title=Quality Evaluator Subagent Invocation,
    colback=purple!5!white,
    colframe=purple!75!black,
    breakable,
    fonttitle=\bfseries
]
\begin{verbatim}
Task(
    subagent_type="survey-quality-evaluator",
    description="Evaluate survey quality",
    prompt="""
    Evaluate the generated survey using a comprehensive 12-dimensional framework.
    
    The survey was generated from [N] papers on [TOPIC], organized into [K] clusters.
    
    EVALUATION FRAMEWORK:
    
    CORE QUALITY (60% weight):
    1. Citation Coverage - % of papers cited effectively
       - Calculate exact percentage of corpus cited
       - Assess distribution across clusters
       - Check for key papers inclusion
    
    2. Accuracy - Factual correctness and attribution
       - Verify claims are properly supported
       - Check author/year attribution accuracy
       - Identify any unsupported generalizations
    
    3. Synthesis Quality - Integration vs mere listing
       - Measure synthesis ratio (integrated vs sequential)
       - Identify cross-paper connections
       - Evaluate comparative analysis depth
    
    4. Organization - Logical flow and structure
       - Assess section/subsection hierarchy
       - Evaluate transition quality
       - Check information progression logic
    
    WRITING QUALITY (20% weight):
    5. Readability - Clarity for target audience
       - Sentence complexity and variety
       - Technical term introduction/explanation
       - Paragraph coherence
    
    6. Academic Rigor - Adherence to scholarly standards
       - Citation format consistency
       - Methodological transparency
       - Limitation acknowledgment
    
    7. Clarity - Precision in technical descriptions
       - Concept explanation quality
       - Ambiguity identification
       - Example usage effectiveness
    
    8. Coherence - Internal consistency
       - Thematic consistency
       - Cross-reference accuracy
       - Narrative flow maintenance
    
    CONTENT DEPTH (20% weight):
    9. Comprehensiveness - Breadth of topic coverage
       - Cluster representation completeness
       - Temporal coverage (publication years)
       - Geographic/institutional diversity
    
    10. Critical Analysis - Depth of evaluation
        - Limitation discussion depth
        - Conflicting findings acknowledgment
        - Methodological critique presence
    
    11. Novelty & Insights - Original contributions
        - Novel connections identified
        - Pattern recognition quality
        - Taxonomy/framework contributions
    
    12. Future Directions - Research trajectory identification
        - Specificity of proposed directions
        - Feasibility assessment
        - Gap identification quality
    
    OUTPUT REQUIREMENTS:
    
    Provide detailed JSON with:
    - dimensional_scores: {
        dimension_name: {
          score: 0-10,
          weight: percentage,
          justification: detailed explanation,
          metrics: quantitative measures,
          specific_examples: 3+ concrete examples
        }
      }
    
    - overall_assessment: {
        weighted_total_score: calculated score,
        score_breakdown: by category,
        quality_level: grade and description,
        publication_readiness: specific assessment
      }
    
    - comparison_to_standards: {
        vs_acm_computing_surveys: assessment,
        vs_conference_surveys: assessment,
        vs_workshop_papers: assessment
      }
    
    - strengths: [7+ specific strengths with evidence]
    
    - weaknesses: [7+ specific weaknesses with evidence]
    
    - prioritized_recommendations: [
        {priority: HIGH/MEDIUM/LOW,
         recommendation: specific action,
         impact: expected improvement,
         effort: implementation difficulty}
      ]
    
    - executive_summary: 200-word synthesis
    
    Use nuanced, context-aware evaluation rather than rigid rules.
    Consider the specific domain, corpus size, and clustering quality.
    Be critical but fair, acknowledging both achievements and gaps.
    
    Save detailed results to: [output_path]/enhanced_evaluation_v3.json
    """
)
\end{verbatim}
\end{tcolorbox}

\end{document}